# Chemical diversity of the atmospheres and interiors of sub-Neptunes: a case study of GJ 436 b

Andrea Guzmán-Mesa[1,2]★ Daniel Kitzmann[1] Christoph Mordasini[2] and Kevin Heng[1,3,4]

[1]*Center for Space and Habitability, Universität Bern, Gesellschaftsstrasse 6, CH-3012 Bern, Switzerland*
[2]*Physikalisches Institut, Universität Bern, Sidlerstrasse 5, CH-3012 Bern, Switzerland*
[3]*University of Warwick, Department of Physics, Astronomy & Astrophysics Group, Coventry CV4 7AL, UK*
[4]*Universitäts-Sternwarte München, Ludwig-Maximilians Universität, Scheinerstraße 1, D-81679 München, Germany*

## ABSTRACT

The atmospheres of sub-Neptunes are expected to exhibit considerable chemical diversity, beyond what is anticipated for gas-giant exoplanets. In the current study, we construct self-consistent radiative transfer and equilibrium chemistry models to explore this chemical diversity. We use GJ 436 b as a case study to further study joint atmosphere-interior models. In particular, we constrain the properties of the interior and atmosphere of the planet based on the available *Spitzer* measurements. While it is possible to fit the emission spectrum of GJ 436 b using a high-metallicity model, we demonstrate that such an atmosphere is inconsistent with physically plausible interior structures. It remains the case that no existing study can adequately fit the 4.5-μm *Spitzer* secondary eclipse measurement, which is probably caused by chemical disequilibrium. Finally, an information content analysis reveals that emission and transmission spectra constrain the carbon-to-oxygen ratio and metallicity at different wavelengths, but the former are less susceptible to flat spectra stemming from highly metal-enriched atmospheres. With the recently launched *James Webb Space Telescope*, we recommend that future analysis of emission and transmission spectra of sub-Neptune planets are carried out self-consistently using both the atmospheric and interior structure models.

## 1 INTRODUCTION

### 1.1 Observational motivation: sub-Neptunes are a common outcome of exoplanet formation

One of the key findings of the NASA's *Kepler* survey is the discovery of a large population of planets with radii between ∼1 and $4R_\oplus$ (Fressin et al. 2013; Petigura, Howard & Marcy 2013) spanning a diverse range of orbital distances and therefore temperatures. The existence of this population was already previously revealed by high-precision radial velocity surveys (Mayor et al. 2011). Exoplanets with radii larger than about 1.6–1.7$R_\oplus$ retain hydrogen envelopes (Rogers 2015), and are commonly referred as 'sub-Neptunes', while those with smaller radii are assumed to be predominantly rocky and are referred as 'super-Earths'. Observations indicate that the populations of super-Earths and sub-Neptunes are separated by an underdensity of planets with radii near 1.7$R_\oplus$ (Fulton et al. 2017; van Eylen et al. 2018). This gap, or valley, is believed to be sculpted by evaporative loss of the hydrogen envelopes of the smallest sub-Neptunes either due to photoevaporation (Owen & Wu 2013; Jin et al. 2014) or core-powered mass-loss (Ginzburg, Schlichting & Sari 2018). In these two scenarios, the position of the valley is well reproduced assuming that the naked-cores are rocky in composition (Jin & Mordasini 2018). Although these two scenarios reproduce well the position of the valley, observations seem to favour photoevaporation. The answer might lie in the dependence on stellar mass. Photoevaporation predicts that the population of sub-Neptunes should shift to lower incident stellar flux with decreasing stellar mass as observed, whereas core powered mass-loss does not depend on stellar mass. For a detailed discussion on the nature and origins of sub-Neptune planets see the review by Bean, Raymond & Owen (2020).

### 1.2 Theoretical motivation: sub-Neptunian atmospheres may be neither purely primary nor secondary

The Solar system provides multiple examples of two different classes of atmospheres: primary and secondary. Jupiter, Saturn, Uranus, and Neptune are examples of primary atmospheres, which are hydrogen-dominated and retain the direct memory or imprints of their formation history (Oberg, Murray-Clay & Bergin 2011; Mordasini et al. 2016). Venus, Mars, and Earth have secondary atmospheres, where the chemistry is mainly controlled by interactions between the atmosphere and interior of the planet. It is plausible that sub-Neptunes, which occupy a continuum between gas/ice giants and terrestrial exoplanets, have hybrid atmospheres that are in between primary and secondary (Kite et al. 2019, 2020). The true composition of sub-Neptunes, however, remains an open question. Zeng et al.

★ E-mail: andrea.guzmanmesa@unibe.ch

(2019) suggested that planets above the $1.7R_\oplus$ threshold are 'water worlds' that have formed beyond the snow line where water and ice are abundant. Venturini et al. (2020a, 2020b) are able to reproduce the two distinct planet populations using a mixture of rocky and lower density water-rich cores.

### 1.3 Theoretical motivation: sub-Neptunian atmospheres may be more chemically diverse

Sub-Neptune atmospheres span a continuum between the relatively well-studied hydrogen-dominated gas giants and the geochemically outgassed atmospheres of rocky exoplanets. Theory predicts a rich chemical diversity of atmospheres for this class of planets and therefore for their observable spectra, beyond what is expected for Jupiter-sized exoplanets (Moses et al. 2013; Hu & Seager 2014). There exists indirect evidence that low-mass exoplanets are enhanced in metallicity and have lower hydrogen–helium mass fractions (Thorngren et al. 2016). High enrichment levels are also predicted by planet formation theory (Fortney 2013). High metallicities are expected to lead to a very diverse chemical inventory of which there is no equivalent in our Solar system.

It remains unclear to what extent the chemical diversity of sub-Neptunes is driven by global quantities like equilibrium temperature and metallicity (Fortney, Visscher & Marley 2020). Whether or not the chemical scenarios predicted by theory truly exist in nature still remains unclear and therefore the understanding of the atmospheres of sub-Neptunes is still primarily based on theoretical expectations.

### 1.4 Case study GJ 436 b

The atmosphere of the warm-Neptune GJ 436 b (Butler et al. 2004; Gillon et al. 2007b) is one of the best-characterized atmospheres to date and at the same time, one of the most challenging to interpret. Besides the measured radius and mass, most of our knowledge about GJ 436 b's atmosphere is based on the *Spitzer* secondary eclipse data published by Stevenson et al. (2010). The measurement consists of photometric secondary eclipses in six *Spitzer* channels from 3.6 to 24 $\mu$m. The original data were reanalyzed by Lanotte, Gillon & Demory (2014). They obtained revised values for the secondary eclipses, with smaller values in the 3.6- and 8-$\mu$m *Spitzer* channels and a deeper 4.5-$\mu$m secondary eclipse compared to Stevenson et al. (2010).

Pont et al. (2009) used NICMOS on the *Hubble Space Telescope* (*HST*) to obtain another transmission spectrum of GJ 436 b. Due to systematic effects, they were unable to detect an expected water absorption signal in their data. A re-analysis of the NICMOS data by Gibson, Pont & Aigrain (2011) found no evidence of other strong molecular absorption features, including water. Beaulieu et al. (2011) analyzed seven primary-transits taken with the *Spitzer* finding that atmosphere of GJ 436 b is dominated mostly by $CH_4$ and $H_2$, with temperature inversion. Knutson et al. (2014a) measured a WFC3 transmission spectrum from 1.1 to 1.7 $\mu$m. Based on their results, they rule out a cloud-free H/He-dominated atmosphere.

Madhusudhan & Seager (2011) used an atmospheric retrieval model to find the best-fitting solution for the secondary eclipse data of GJ 436 b. They find that the best-fitting consists of an atmosphere rich in CO and $CO_2$ but poor in $CH_4$. As discussed in Moses et al. (2013), such an atmosphere could be the natural consequence of a very high atmospheric metallicity. An alternative possibility to obtain such an outcome is presented in Hu, Seager & Yung (2015). They propose the existence of a helium-dominated atmosphere where, due to the lack of hydrogen, the main carrier of carbon would thus be CO and $CO_2$ rather than $CH_4$.

Morley et al. (2017) analyzed the *Spitzer* secondary eclipse observations together with the *HST/WFC3* transmission spectrum from Knutson et al. (2014a) with self-consistent atmospheric models and retrieval codes. In general, their best-fitting solutions have very high metallicities of about 1000 × solar but fail to fit all measured *Spitzer* photometry points. Most notably, their models underpredict the 15-$\mu$m channel, while overestimating the 4.5-$\mu$m one. Adding the impact of disequilibrium chemistry via quenching approximation or the effect of clouds to their model did not provide a better overall fit.

Lewis et al. (2010) studied how atmospheric metallicity influences general circulation models applied to GJ 436 b. They find that models with high metallicity and disequilibrium chemistry are favoured. Line et al. (2011) implemented a model which includes vertical mixing, photochemistry and chemical kinetics finding that methane should be the predominant carbon carrier/reservoir in GJ 436 b's atmosphere.

The high flux at 3.6 $\mu$m together with the low flux at 4.5 $\mu$m indicate that CO and not $CH_4$ is the dominant carbon carrier by orders of magnitude in GJ 436 b's atmosphere. This is in contrast to what chemical equilibrium models predict. So far, none of the atmospheric models of GJ 436 b seem to provide a fully consistent explanation to the transit and secondary-eclipse data. However, there seems to be a suggestion that high metallicity, disequilibrium chemistry and efficient tidal heating (Agúndez et al. 2014; Morley et al. 2017) could be at play.

### 1.5 Goals and layout of the current study

In this study, we calculate fully self-consistent coupled atmospheric and interior models of sub-Neptunes assuming equilibrium chemistry. These calculations are post-processed to predict emission and transmission spectra, which we further analyse for their information content. As a case study, we focus on GJ 436 b for which we compute a grid of models and use machine learning methods to interpret its low-resolution spectrophotometry.

In Sections 2 and 3, we describe our methodology involving radiative transfer, opacities, ray-tracing, equilibrium chemistry and interior structure. In Section 4, we present the results of our coupled atmospheric and interior models, as well as their application to GJ 436 b. In Section 5, we discuss the observational implications of the chemical diversity of sub-Neptunes.

## 2 CHEMICAL DIVERSITY OF SUB-NEPTUNE ATMOSPHERES

In this section, we first explore the potential chemical diversity of sub-Neptune atmospheres. This part is guided by the work of Moses et al. (2013), who already provide estimates for the compositional diversity that could be expected for these types of atmospheres. However, in contrast to their study which uses only a chemistry model, we here employ self-consistent atmospheric model calculations that yield the vertical atmospheric structure. We also present the corresponding emission and transmission spectra for the computed planetary atmospheres. To model a typical sub-Neptune with an equilibrium temperature of roughly 955 K, we choose the following set of parameters: $\log g = 1000$ cm s$^{-2}$, orbital distance $a = 0.015$ au, $R_p = 0.27 R_{Jup}$, $R_\star = 0.34 R_\odot$, and $T_\star = 3500$K.

### 2.1 Model description

In the following, we briefly summarize the coupled atmosphere–chemistry model. In particular, we use the HELIOS atmosphere model together with the equilibrium chemistry FastChem and the absorption cross-sections of molecules and atoms provided by our `HELIOS-K` opacity calculator. In a post-process, we use our `Helios-O` model to derive the transmission spectra from the computed atmospheric structures.

### 2.1.1 HELIOS atmospheric model

We compute the atmosphere's thermal structure and emission spectrum in radiative-convective equilibrium using `HELIOS` (Malik et al. 2017, 2019), which is an open-source, 1D atmospheric model. It uses a hemispheric two-stream method with an improved treatment of scattering (Heng & Kitzmann 2017; Heng, Matej & Kitzmann 2018). `HELIOS` is a self-consistent atmospheric model, meaning that the thermal structure is an outcome of the computation, rather than an underlying assumption.

The radiative transfer calculations are performed using the opacity sampling method (see e.g. Malik et al. 2017). They cover a wavelength range from 0.3 to 200 µm with a constant spectral resolution of $\lambda/\Delta\lambda = 3000$. We employ an Eddington coefficient $\epsilon_2$ of 2/3 (see Malik et al. 2019). Convective adjustment is performed using the dry adiabatic lapse rate, employing the Schwarzschild criterion to check for convectively unstable regions within the atmospheric column. We assume an efficient global heat circulation and therefore set the heat redistribution factor *f* to 0.25 (Malik et al. 2017). The pressure at the bottom of the computed atmosphere is set to 1 kbar for all cases.

### 2.1.2 HELIOS-K opacity calculator

One of the key ingredients of modelling and interpreting spectra of exoplanet atmospheres is the computation of opacities. `HELIOS-K` is an ultrafast, open-source, GPU-accelerated opacity calculator that takes line lists from the `ExoMol`, `HITRAN`, `HITEMP`, `NIST`, `Kurucz`, and `VALD3` databases and computes the opacity function of the atmosphere for any combination of atoms and molecules (Grimm & Heng 2015; Grimm et al. 2021).

In the current work, we consider the following molecular species: $H_2O$ (Barber et al. 2006; Polyansky et al. 2018), CO (Li et al. 2015), $NH_3$ (Yurchenko, Barber & Tennyson 2011), $CH_4$ (Yurchenko et al. 2013; Yurchenko & Tennyson 2014), HCN (Harris et al. 2006; Barber et al. 2014), $CO_2$ (Rothman et al. 2010), $C_2H_2$ (Gordon, Rothman & Hill 2016), and $O_3$ (Rothman et al. 2013). Their line lists are obtained from the `ExoMol`, `HITRAN` (Rothman et al. 1987, 1992, 1998, 2003, 2005, 2009, 2013) and `HITEMP` (Rothman et al. 2010) spectroscopic databases. Opacities are calculated with a constant wavenumber step of 0.1 cm$^{-1}$ for wavenumbers between 40 000 and 0 cm$^{-1}$. We use a Voigt profile with a line wing cutoff at 100 cm$^{-1}$ from the line center. Pressure broadening and self-broadening parameters are used as provided by default in the `ExoMol` and `HITRAN` online spectroscopic databases.

In addition, we also include collision-induced absorption (CIA) opacities of $H_2$–$H_2$ and $H_2$–He (Abel et al. 2011; Richard et al. 2012).

### 2.1.3 FastChem chemical equilibrium calculator

To calculate the chemical-equilibrium volume mixing ratios, we use the open-source equilibrium chemistry calculator `FastChem`, which is based on a semi-analytical approach as outlined in Stock et al. (2018). `FastChem` includes more than 550 gas-phase chemical species, including ions, for elements more abundant than germanium. Here, we compute the thermochemical equilibrium atmospheric abundances for temperatures between 100–6000 K and pressures from $10^{-13}$ bar to 1000 bar. For our work, we assume solar elemental abundances as detailed in table 1 of Asplund et al. (2009). As input, we use the temperature–pressure profiles generated self-consistently with `HELIOS`.

We use the C/H ratio as a general scaling factor for the overall metallicity M/H, i.e. all elements except for H, He, and O are scaled by this factor (Heng 2018). When varying the metallicity, we increase the abundances of all elements, except those of H, He, and Ne, following the approach of Moses et al. (2013). Since all elements are scaled by the same factor, the ratios of their element abundances remains solar. The only exception is the carbon-to-oxygen ratio (C/O), which is used as another free parameter. Here, we scale the carbon abundance with the assumed metallicity factor and compute the oxygen element abundance with the given C/O ratio. Condensates are currently not considered in our `FastChem` calculations.

### 2.1.4 Helios-O ray-tracing code for transmission spectra (post-processing)

Upon obtaining the self-consistent thermal structure of a model atmosphere, one may perform post-processing to obtain the transmission spectrum. For this purpose, we use the ray-tracing code `Helios-O`, which computes wavelength-dependent transit chords to construct the transmission spectrum (Gaidos, Kitzmann & Heng 2017; Bower et al. 2019). The transmission spectra are normalized to the chosen white-light radius of the planet.

## 2.2 Atmospheric models: composition and spectra

In Fig. 1, we show several examples of simulated atmospheres of sub-Neptunes, where both the C/O and elemental carbon abundance (C/H) are allowed to vary (see Section 2.1.3 for details). These plots are inspired by fig. 4 of Moses et al. (2013). However, we derive them from self-consistent radiative transfer and equilibrium chemistry models, rather than assume fixed values of temperature and pressure.

Generally, the chemical composition of an atmosphere varies throughout the atmosphere. For the pie charts in Fig. 1, we present the mixing ratios for the chemical species at pressures where the contribution functions of the radiation flux peak for visible wavelengths. This typically corresponds to pressures near 0.1 bar in many cases but can also increase to values of about 10 bar for, e.g. C/H=300 × solar and C/O ratio of 0.1 (not shown). Thus, these plots represent the chemical composition at the planet's respective photosphere. In the appendix, we additionally show the vertical, chemical abundances of selected species (see Figs A1–A4), as well as the temperature–pressure profiles (Fig. A5), for completeness.

The atmospheric composition of sub-Neptunes has a strong dependence on the atmospheric properties and assumed elemental abundances. For example, the influence of C/O on the atmospheric chemistry has been well studied (e.g. Madhusudhan 2012). In general, at high metallicities, they can become either CO- and $CO_2$-dominated or $O_2$-rich atmospheres, depending on their C/O ratio. On the other hand, at low-metallicities hydrogen-dominated atmospheres are the prevalent outcome.

The two top rows of Fig. 1 show that for solar-like C/O ratios, $H_2$ becomes less abundant with increasing metallicity, giving way to other species like $H_2O$, $CO_2$, CO, and $N_2$, as expected. For an atmosphere with super-solar elemental abundances (second row), the chemical composition is mostly dominated by $H_2O$ at low C/O ratios, resulting in a 'water world' (Moses et al. 2013). With increasing carbon-to-oxygen ratios, the $H_2O$ abundances start to decrease, resulting in increased mixing ratios of other hydrogen and oxygen-carriers, like CO or $CH_4$, for example. These results are consistent with those presented in Moses et al. (2013).

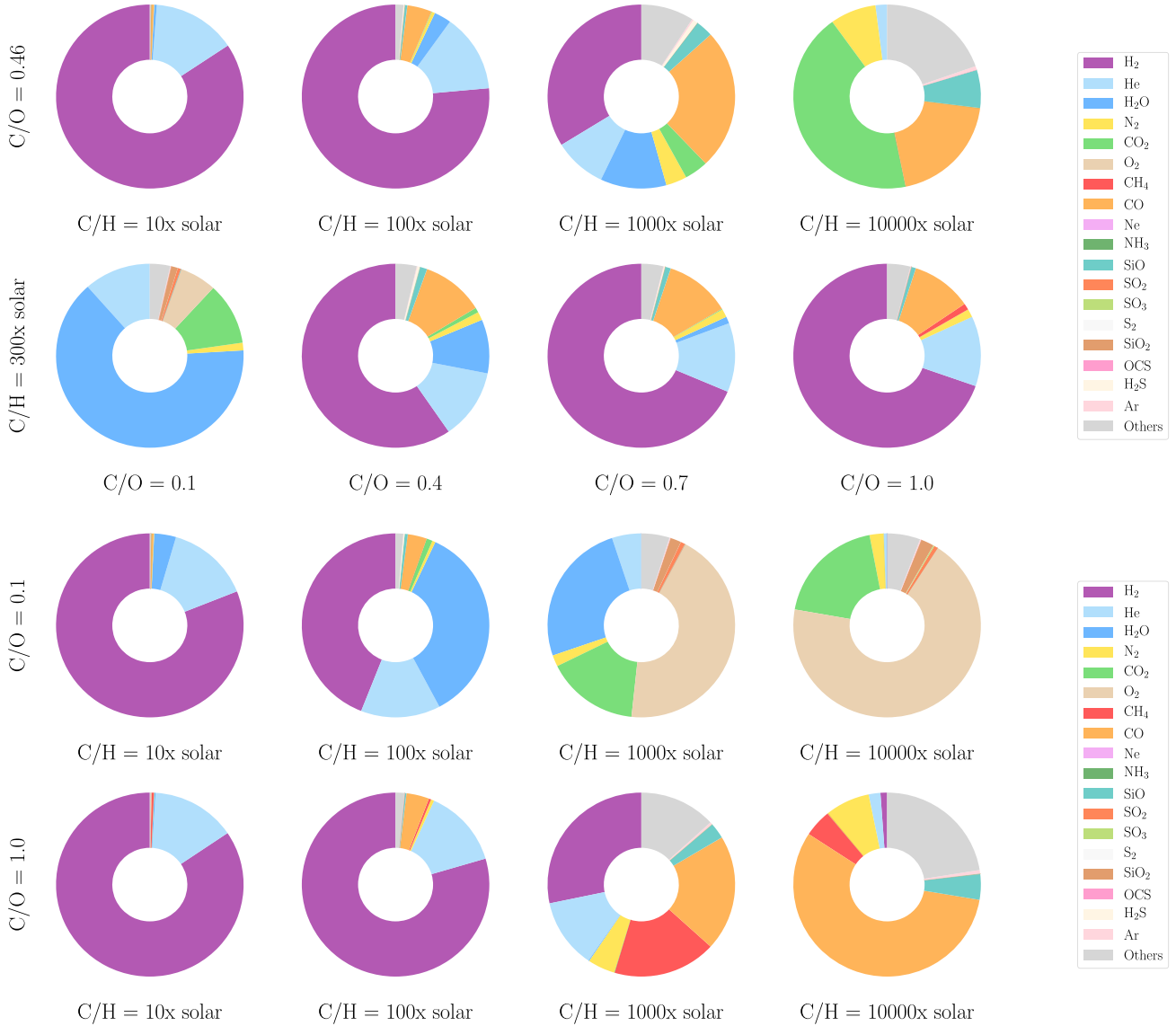

**Figure 1.** Theory predicts chemical diversity in the atmospheres of sub-Neptunes beyond what is expected for gas-giant exoplanets. These 'chemical pie charts' represent the volume mixing ratios of the different atmospheric constituents. They are motivated by the work of Moses et al. (2013), but are derived from self-consistent radiative transfer and equilibrium chemistry models. The carbon-to-oxygen ratio (C/O) and elemental carbon abundance (C/H) of the top two rows are varied according to fig. 4 of Moses et al. (2013). In the two bottom rows, we repeat the calculations of the top one but for C/O = 0.1 and 1. The self-consistent temperature–pressure profiles corresponding to these different scenarios are shown in Fig. A5. The pressures at which these compositions are shown is taken from the peak of the contribution functions at visible wavelengths.

In the third row of Fig. 1, we show the variation of the chemical composition for a fixed sub-solar C/O ratio of 0.1 with the overall metallicity. Due to the lack of the carbon, oxygen-bearing species like water are dominating the chemical composition as the metallicity increases beyond 10 × solar. For the highest assumed C/H ratios, the atmospheres become essentially largely dominated by $O_2$.

On the other hand, atmospheres with a carbon-to-oxygen ratio of unity (bottom row of Fig. 1) are distinctly different. Here, the carbon becomes essentially locked in CO and $CH_4$ (Madhusudhan 2012). The oxygen is mostly bound in CO, as well as multiple other oxygen-bearing species, like SiO. At very high C/H ratios, the atmosphere is dominated by carbon monoxide to a large degree, with only minor contributions by $CH_4$ as a carbon carrier. Due to the very strong chemical stability of the CO molecule, the majority of the C and O atoms are locked in this molecule. Additionally, the abundance of molecular nitrogen also increases with the overall metallicity, but does never dominate the atmospheric composition for any case. Abundance profiles that are uniform with height or pressure are unaffected by chemical disequilibrium driven by vertical atmospheric mixing (quenching) (see Figs A1, A2, A3 and A4). For chemical species with such profiles, chemical disequilibrium may instead be driven by photochemistry.

In Fig. 2, we present the corresponding emission (left-hand panel) and transmission spectra (right-hand panel) for the different cases. At very high metallicities, the mean molecular weight of the atmosphere is so high that the overall atmospheric scale heights are decreasing quite strongly. For these cases, therefore, the transmission spectra are more or less flat. Such an atmosphere would be difficult to observe via transmission spectroscopy. As Fig. 2 suggests, emission spectra are less affected by the mean molecular weight. Thus, such atmospheres, if they exist, should be observed in emission, rather than in transmission, if observationally feasible.

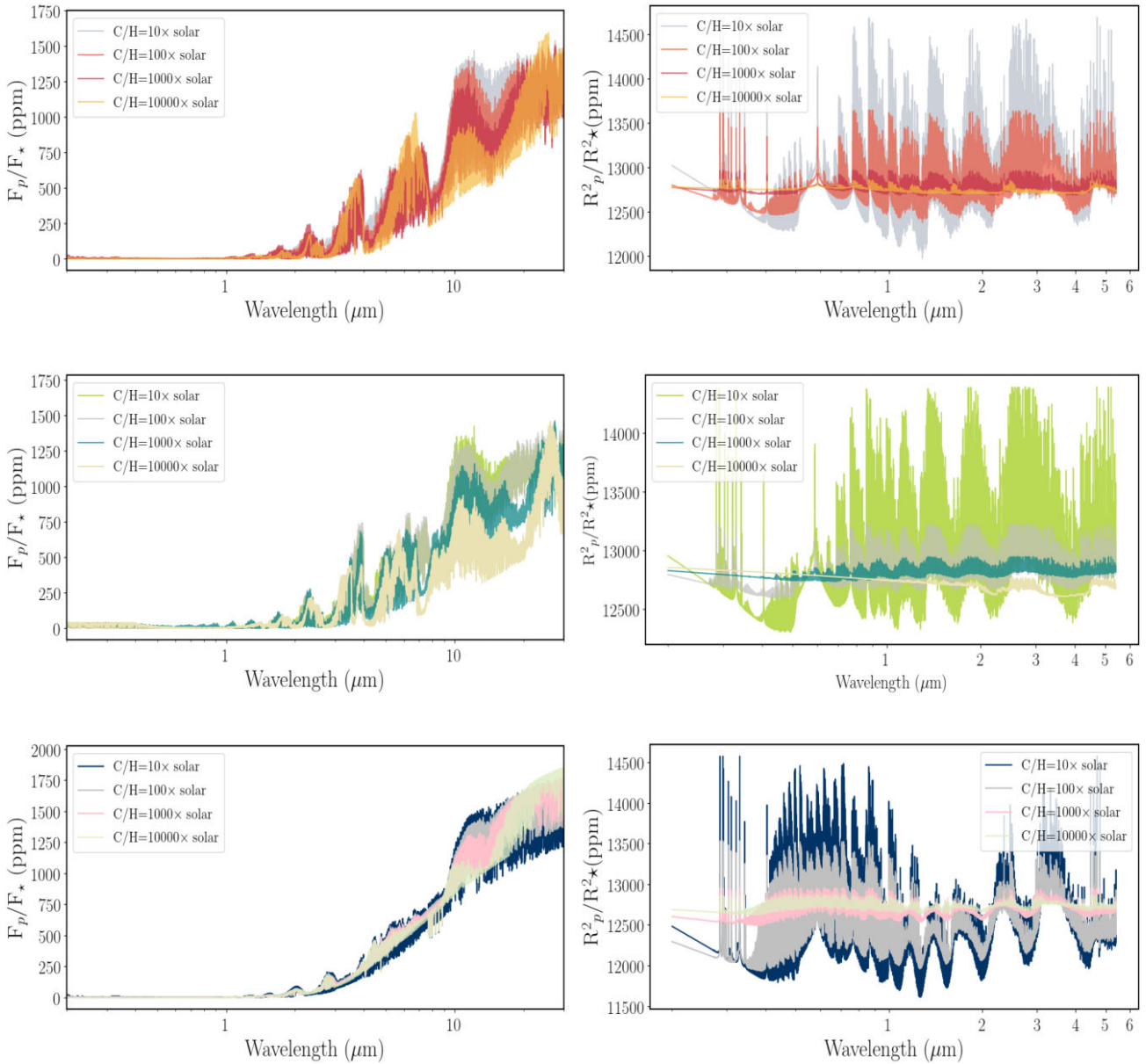

**Figure 2.** Each row shows the emission spectra (left-hand panel) and its corresponding transmission spectra (right-hand panel) for different values of C/H at a fixed carbon-to-oxygen ratio. The top row has a solar C/O ratio, the middle row has C/O = 0.1 and the bottom row C/O = 1.0. For consistency, we have used the same colour schemes for each corresponding pair of emission and transmission spectra. These spectra correspond to the first, third, and fourth rows of Fig. 1.

## 3 SELF-CONSISTENT COUPLING OF INTERIOR AND ATMOSPHERIC MODELS OF SUB-NEPTUNES

Linking the atmospheric properties to planetary interior structure models can provide valuable insights into the interior of sub-Neptune planets, namely their bulk composition. It can also place addi-tional independent constraints on possible atmospheric compositions (Thorngren & Fortney 2019). We now extend the work of the previous section by coupling the atmospheric model to a model of the interior.

For the description of the planet's interior, we use the planetary interior structure and evolution model `Completo21` (Mordasini et al. 2012b; Jin et al. 2014) to couple the planet's interior to the atmospheric model given by `HELIOS` (Malik et al. 2017, 2019). Similar work employing these codes has already been done by Linder et al. (2019) and Marleau et al. (2019).

### 3.1 Completo21 planetary interior structure code

`Completo21` solves the classical 1D spherically symmetric planetary interior structure equations (e.g. Bodenheimer & Pollack 1986).

This yields the pressure, temperature, enclosed mass, and intrinsic luminosity as a function of distance from the planetary centre. To close this system of coupled partial differential equations, boundary conditions and equations of state (EOS) are needed. Completo21 models the planets as consisting of three different layers: an iron core, a silicate mantle, and a gaseous/fluid envelope. An Earth-like 2:1 silicate:iron ratio is assumed in all cases.

The silicate and iron parts are described by the modified polytropic equation of state by Seager et al. (2007), using the parameters of perovskite for the silicates. The gaseous/fluid envelope is assumed to be made of a homogeneous mixture of H/He described by the new CMS equation of state from Chabrier, Mazevet & Soubiran (2019) and of $H_2O$ described by the AQUA equation of state (Haldemann et al. 2020). The mass fraction of water is denoted by $Z$, the rest being H/He. The H/He and $H_2O$ are combined according to the additive volume law (Baraffe, Chabrier & Barman 2008). In contrast to the EOS used in the atmosphere, these are non-ideal equations of state that are necessary to model the very high pressures and densities in the interior.

H, He, and $H_2O$ are often among the species that make up dominant parts of the atmospheric chemical composition (Fig. 1). This is in particular the case at low to intermediate C/H and/or low C/O. Thus, it makes sense to use an EOS specifically for H/He + $H_2O$ for such cases. However, especially at high C/H and C/O, the composition can also be dominated by other species like CO, $CO_2$, $N_2$, and $O_2$, as also seen in these figures. To the best of our knowledge, there is, however, an absence of non-ideal equations of state for planetary interiors for such general mixtures, making an approximate treatment necessary. Thus, as in Valencia et al. (2013), we use the H/He + $H_2O$ mixture as a representation of all compositions, but determining the water mass fraction $Z$ in a special way as described next. It is clear that the specific thermodynamic properties of water affect the results, but given the absence of general non-ideal EOS, this is the currently viable approximation we have to use.

As described, we assume a fully mixed H/He + water layer instead of a pure $H_2O$ layer or a pure H/He envelope. In the Solar system, there is evidence that suggests that mixing occurs in the diluted core of Jupiter (Wahl et al. 2017). In the case of Uranus and Neptune, the range of compositions allowed (that fit the observed mass and radius, but also the gravitational moments and the atmospheric composition) is very wide. For them, both layered and mixed compositions are possible (Helled, Nettelmann & Guillot 2020). The reality is, however, that there are still substantial uncertainties regarding their bulk compositions and internal structures. This is probably no different for sub-Neptune exoplanets.

Recent work suggests that ice can substantially mix with the silicate layer (Vazan, Sari & Kessel 2022). This is, however, not considered in the current model.

### 3.2 Coupling the atmosphere and interior structures

To calculate an interior structure for a given $\log g$, equilibrium temperature, and chemical composition, the interior structure model also needs outer boundary conditions besides the EOS. These are the pressure, temperature, and total planet mass at the point where the atmosphere (provided by HELIOS) and interior (provided by Completo21) are joined. The intrinsic luminosity at this point, which is calculated from $T_{int}$, is also a boundary condition.

We follow the general approach of Chabrier & Baraffe (1997) to couple atmospheres and interior, and choose a fixed pressure of 1 kbar as the connection point (the so-called bottom of the atmosphere or BOA). The choice of 1 kbar is a compromise between having a sufficiently high pressure to be in the inner convective zone of warm irradiated planets (Guillot & Showman 2002), and a sufficiently low pressure such that the mass contained in the atmosphere, relative to the one in the envelope, is still small.

The temperature at the BOA is a direct outcome of the HELIOS radiative transfer calculations. Via the FastChem chemical composition calculation, we also have the mean molecular weight at this position. In the atmosphere model, which assumes an ideal equation of state, temperature, pressure and mean molecular weight can be used to calculate the density at the BOA. We then use this density to determine for the interior structure model the water mass fraction $Z$ such that the density is continuous. This means that for low C/H and C/O, a correspondingly low $Z$ will result, while high C/H and C/O lead to high $Z$. Note that the density is indirectly also influenced via the temperature by the different optical properties of the molecules.

In Fig. 3, we show the logistical flow of the coupling of Completo21 to HELIOS. A full description of the quantities and the flow is provided in the caption.

#### 3.2.1 Density scaling

In the interior, we assume water as being representative of all species different than H/He. Because of heavy molecules like $CO_2$, this can lead to the situation that the density found in the interior model for a 100 per cent water composition is less than the density obtained in the atmosphere model (with its ideal EOS). This would lead to an unphysical, unstable density inversion (Thorngren & Fortney 2019). In such cases, we determine as an approximation a scaling factor (>1) such that the density is continuous, and then apply this scaling factor uniformly to all further densities calculations in the interior EOS. This situation happens for chemical compositions with high C/O and high C/H where molecules heavier than $H_2O$ (like CO, $CO_2$) lead to higher density than a pure water composition (the highest density possible in the interior EOS). We found, however, that for all converged structures (see Section 3.3), no density scaling was actually necessary.

#### 3.2.2 Rosseland opacity

Care is also taken to not introduce artefacts caused by the different techniques for calculating the Rosseland opacity in the atmosphere and the interior. In the interior model, the Rosseland opacity is used to decide if a layer is convective or radiative. In the interior, the opacity tables of Freedman et al. (2013) for a condensate-free scaled solar-composition composition gas are used. This leads to similar results as the Rosseland opacity obtained directly in the atmosphere model provided that the enrichment level is not too high (i.e. Freedman et al. 2013 extend only to 50 × solar) and that the C/O is not too different from the solar value. In other cases, the opacity obtained in the interior structure model is less realistic than the one in the atmosphere model, and at the BOA, there will, in particular, be a jump in the opacities obtained in the two models. To address this, we proceed in a similar way as for the density: we derive a scaling factor we use to scale the opacity in the interior model such that the opacities are continuous across the BOA, and then apply this scaling factor as an approximation to all opacity calculations in the interior structure model. In this way, it is found that the BOA always lies in the internal convective zone, as predicted by both models. Since for convective layers, the opacity does not enter the structure equations (in particular the temperature gradient), the specific opacities of the interior structure model are actually not used at all in the structure calculations.

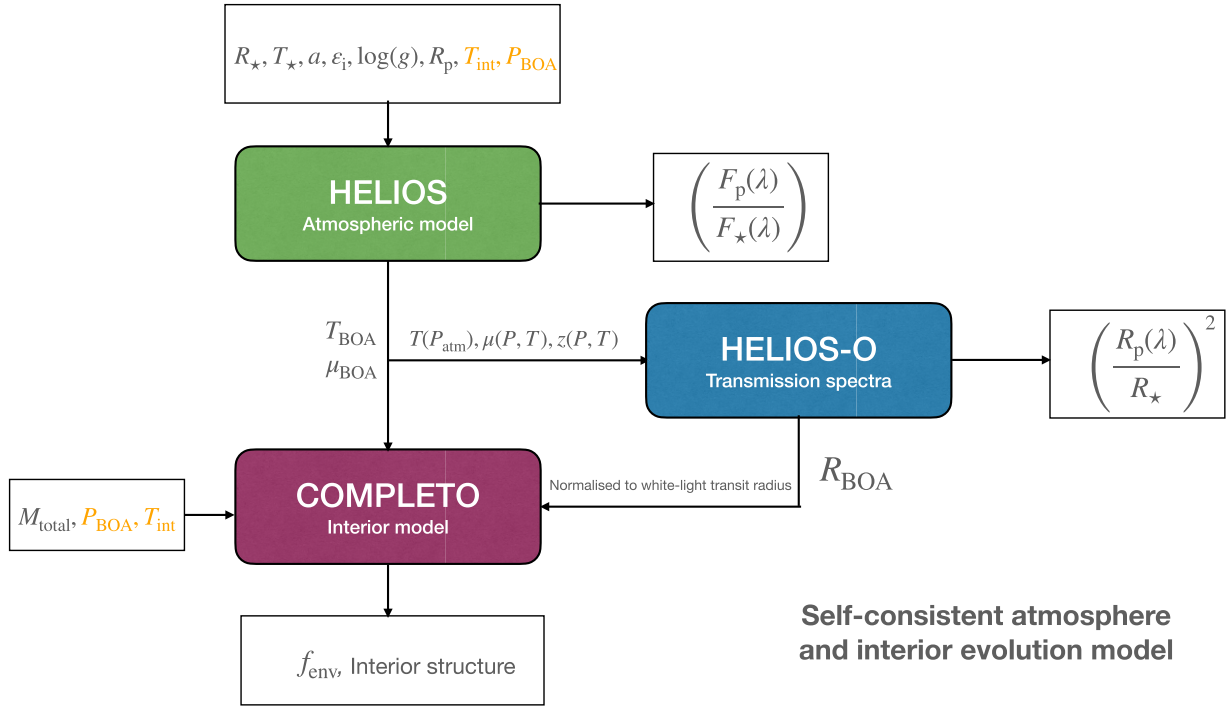

**Figure 3.** Schematic of the logistical flow of HELIOS versus Completo21. HELIOS takes as an input the radius $R_\star$ and temperature $T_\star$ of the star, the star–planet separation $a$, the elemental abundances $\varepsilon_i$, the pressure at the bottom of the atmosphere $P_{BOA}$, the interior temperature of the planet $T_{int}$, the planet's surface gravity $\log g$ and the radius of the planet $R_p$. Some of the outputs of HELIOS for the atmosphere are the P-T structure $T(P_{atm})$, the molecular weight $\mu(P, T)$ and the altitude $z(P, T)$, as well as the emission spectra and secondary eclipse depth $(F_p(\lambda)\backslash F_\star(\lambda))$ of the planet. Some of these outcomes are then used in the post-processing stage with Helios-O to compute the normalized transmission spectra $(R_p(\lambda)\backslash R_\star)^2$ from which we also obtained the planet radius at the bottom of the atmosphere $R_{BOA}$ used as an input in Completo21. Completo21 takes as an input the same $T_{int}$ and $P_{BOA}$ (indicated in yellow) used in HELIOS. In addition, it also receives the total mass of the planet $M_{total}$, the temperature $T_{BOA}$ and the molecular weight $\mu_{BOA}$ at the bottom of the atmospheres from the HELIOS outputs. Completo21 provides the envelope mass fraction $f_{env}$ and the interior structure of the planet as an outcome.

*3.2.3 Determination of the envelope mass fraction*

As a last constraint, we get from Helios-O the radius $R_{BOA}$ at which the BOA must be situated in order to reproduce the correct observed transit radius. This quantity is obtained by normalizing the calculated transmission spectra to the measured white-light radius of the considered planet. This thickness of the atmosphere depends in particular on the atmospheric composition. We use this observational constraint to obtain a key quantity describing the planet interior which is at this point not yet known: the envelope mass fraction $f_{env}$. With the observationally given total mass, this then yields the mass of the envelope and of the silicate/iron core.

To find $f_{env}$, we proceed iteratively: for a trial $f_{env}$, we calculate the interior structure. If the trial $f_{env}$ was too high, the numerically found radius at 1 kbar will be higher than the $R_{BOA}$ required to fit the observed transit radius. Thus, for the next iteration, $f_{env}$ is reduced until a converged solution (i.e. reproducing both observed mass and radius, for the given assumed composition) is reached.

Note that such a converged solution might not be found for any assumed atmospheric composition: for example, for a very high C/H and C/O composition, the gas density can be so high that it becomes impossible to fit the observed (lower) mean density of a planet, even in the complete absence of an even denser silicate-iron core. The corresponding C/H and C/O pair (which might be allowed from spectral constraints alone) is then excluded on grounds of the observed mean planet density. This means that the interior structure calculation adds an additional constraint on the allowed atmospheric composition which is independent of the spectrum (under the assumption of homogeneous mixing, of course).

**3.3 Case study: joint atmosphere-interior models for the sub-Neptune GJ 436 b**

In this subsection, we present the outcome of a coupled model calculation for the sub-Neptune GJ 436 b as an example. For these calculations, we use the general parameters listed in Table 1. We use a metallicity of C/H = 100 × solar and a C/O ratio of 0.5 similar to that of the C/H = 100 × solar and C/O = 0.46 case, which is depicted in the top row of Fig. 1. This atmosphere is made of mostly $H_2$ and CO with some significant contributions from He, $H_2O$, $CO_2$, and $N_2$. For this specific case, $M_{core} = 18.62 M_\oplus$, $M_{env} = 6.85 M_\oplus$, and $Z = 0.62$. The results for the coupled model are presented in Fig. 4.

In the top left-hand panel of Fig. 4, we show the pressure–temperature profile. The parts modelled by the atmospheric and interior structure codes are show with different colours. The interior structure of the iron-silicate core is not shown, but is assumed to have an adiabatic temperature gradient. We see that as we go deeper into the planet, the temperature increases as expected up to ∼14 000K at the core-envelope boundary (right end of the red line). For comparison, the temperature at the core-envelope interface for Jupiter is estimated to be around 20 000 to 30 000 K (Debras & Chabrier 2019). One notes the slightly different temperature gradients at the transition from `HELIOS` and `Completo21`, which is a consequence of the different EOS that are used. The radiative-convective boundary (RCB) occurs at a pressure of ∼100 bar, i.e. inside of the atmosphere, as it should be for the coupling.

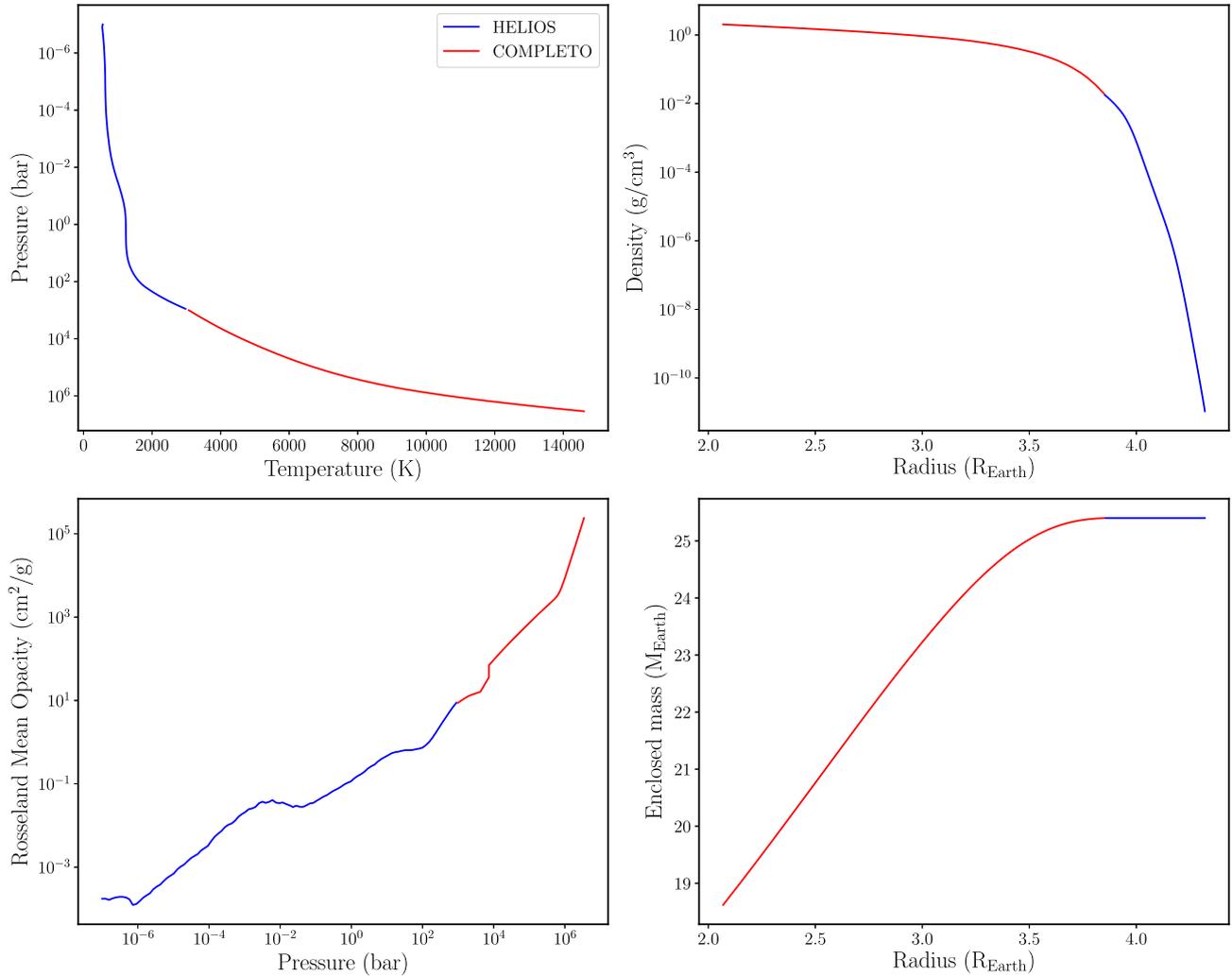

**Figure 4.** The four panels show the coupling of the atmosphere model `HELIOS` (blue) and the interior model `Completo21` (red) for one of the grid scenarios for GJ 436 b. This case corresponds to C/H=100 × solar and C/O = 0.5. The resulting core and envelope mass are 18.62 and 6.85$M_\oplus$, respectively. The structure of the silicate-iron core is not shown. The panels show the pressure-temperature diagram (top left-hand panel), the density as a function of radius (distance from the planet's centre, top right-hand panel), the Rosseland mean opacity as a function of pressure (bottom left-hand panel), and the enclosed mass as a function of radius.

**Table 1.** GJ 436 stellar and planetary parameters.

| Parameter | Value | Reference |
|---|---|---|
| Stellar | | |
| $R_\star$ | 0.449$R_\odot$ | Bourrier et al. (2018) |
| $T_\star$ | 3660 K | Gaia Collaboration (2018) |
| Planet | | |
| Surface $g$ | 1318 cm/s$^2$ | von Braun et al. (2012) |
| $R_p$ | 0.3739 $R_J$ | Knutson et al. (2014a), Turner et al. (2016) |
| $M_p$ | 25.4$M_\oplus$ | Bourrier et al. (2018) |
| $a$ | 0.028 au | Triffonov et al. (2018) |
| $\rho$ | 1.73 g/cm$^3$ | Turner et al. (2016) |

In the top right-hand panel, we show the density profile as a function of the planet's radius. The core radius is about 2.1$R_\oplus$, while the total radius (including the outer very tenuous layers) is 4.32$R_\oplus$. This value is, as expected, a bit larger than the transit radius of 4.17$R_\oplus$.

The density is continuous and monotonically decreasing with increasing distance from the planet's core. As in all converged interior structure models (see Section 4.4), no artificial scaling of the density was necessary (see Section 3.2.1) to make it continuous across the atmosphere-interior boundary. In the inner part, the density varies by about one and a half orders of magnitudes. The density at the core-envelope interface is ∼ 2.03 g cm$^{-3}$. This is a fluid-like density, illustrating clearly why also for a planet without a very massive envelope, an ideal equation of state could not be used for the interior. In the outer parts modelled by the atmospheric model, the density decreases in contrast strongly by many orders of magnitudes, making there the ideal EOS applicable.

We also show the Rosseland mean opacity as a function of pressure in the bottom left-hand panel. The Rosseland opacity is by construction continuous (see Section 3.2.2) and increases as expected with pressure and temperature (e.g. Freedman et al. 2013). To make the opacities continuous, a scaling factor of 0.426 was applied to the opacities in the interior model. The opacity is particularly high in the deep atmosphere, and even more so in the interior. This is indicative of the occurrence of convection in the envelope and lower atmosphere, whereas the upper atmospheric temperatures are determined by radiative energy transport. The RCB is as mentioned at a pressure of ∼100 bar, meaning that the specific values of the opacity at even higher pressures do not enter the calculation.

Finally, we show in the bottom right-hand panel the enclosed mass as a function of the radius of the planet. The core mass, in this case, is $18.62 M_\oplus$. As expected from the density structure, we see that almost all of the planet's mass is contained in the interior of the planet, with only a small fraction $0.0088 M_\oplus$ (i.e, 0.03 per cent of the total planet's mass) of it forming the atmosphere. This justifies a posterior our approach to neglect the mass contained in the atmosphere.

Both codes for computing the interior and the atmospheric structure pose some limitations. HELIOS, for example, uses an EOS that assumes an ideal gas. However, this might not hold true for the high-pressure regions of the deeper atmosphere. Furthermore, HELIOS enforces chemical equilibrium via the FastChem chemistry model throughout the entire atmosphere. Since FastChem also assumes an ideal gas, the chemical composition of the lower atmosphere predicted by HELIOS might differ from one with a more accurate equation of state.

Completo21 on the other hand assumes a chemically homogeneous envelope with an adiabatic temperature gradient. The silicate-iron core is assumed to be completely separated from the envelope, and the silicate and iron parts are also each homogeneous. Recent studies of the Solar system giant planets (Nettelmann et al. 2013; Wahl et al. 2017; Debras & Chabrier 2019; Scheibe, Nettelmann & Redmer 2021) show that these planets likely have more complex interiors with compositional gradients, non-convective parts, and fuzzy cores. This would affect our results for, e.g. the inferred envelope mass fraction. The difficulty in applying these more complex models to exoplanets lies in the absence of observational constraints that are only available for the Solar system counterparts (gravitational moments, intrinsic luminosity, and detailed atmospheric composition).

## 4 JOINT ATMOSPHERE-INTERIOR RETRIEVAL ANALYSIS OF GJ 436 B

After the general discussion of the potential chemical diversity in sub-Neptunes, we now focus on a specific observed planet. We use GJ 436 b as an important example of a close-in Neptunian planet, and study it with the joint atmosphere and interior model. In particular, we aim to constrain the properties of the interior and atmosphere of the planet based on the available measurement presented in Lanotte et al. (2014). These measurements include secondary-eclipse and transmission observations in multiple *Spitzer* bandpasses.

To constrain the planet's chemical composition in terms of C/O ratio and the overall metallicity C/H, based on the available observational data, we use the random forest machine learning method. In particular, we use the HELA model that has previously also been used in Márquez-Neila et al. (2018), Fisher et al. (2020), Guzmán-Mesa et al. (2020), Oreshenko et al. (2019) for example.

### 4.1 Available GJ436 data used in this study

HELIOS requires fundamental parameters for both the star and the planet. As an illustration, we fix these parameter values to those of GJ 436 and GJ 436 b, respectively (see Table 1). For GJ 436, we use a radius of $R_\star = 0.449\,R_\odot$ (Bourrier et al. 2018) and a stellar effective temperature of $T_\star = 3660$ K (Gaia Collaboration 2018).

For GJ 436 b, we use a surface gravity of $g = 1318$ cm/s$^{-2}$ (von Braun et al. 2012) and a radius $R_p = 0.3739\,R_J$. The radius is equal to the measured white-light transit radius of GJ 436 b (Knutson et al. 2014a; Turner et al. 2016).

### 4.2 HELA random forest retrieval for exoplanet atmospheres

HELA is a retrieval algorithm that uses the supervised machine learning method of the random forest (Ho 1998; Breiman 2001) to perform atmospheric retrievals on transmission and emission spectra of exoplanets (Márquez-Neila et al. 2018). Two important byproducts of the random forest technique are the 'feature importance' plots, which quantify the relative importance of each data point in the spectrum for constraining each parameter, and the 'real versus predicted' (RvP), which quantify the degree to which each parameter may be predicted in mock retrievals given a noise model. It has been used to perform information content analysis of *James Webb Space Telescope* (*JWST*) transmission spectra of sub-Neptune planets (Guzmán-Mesa et al. 2020).

For the random forest, we require a grid of atmosphere-interior models, together with the corresponding spectra. The grid, as a function of C/O and C/H is computed using the coupled interior-atmosphere model described in the last section, together with the Helios-O code to provide post-processed transmission spectra.

Using the stellar and planetary parameters from Table 1, we compute a grid of cloud-free atmospheric models, spanning a range of values for C/O from 0.1 to 1.0 in intervals of 0.1 and log C/H from 0 to 4 × solar in steps of 0.5. In total, we compute 90 self-consistent atmospheric models. The star–planet separation is $a = 0.028$ au (Triffonov et al. 2018). The internal temperature $T_{int}$ depends in principle on formation and evolution (Mordasini, Marleau & Mollière 2017; Linder et al. 2019) as well as bloating mechanisms (Pu & Valencia 2017; Sarkis et al. 2021). Empirically, the internal temperatures of warm sub-Neptunes are unknown. For illustration, we assume an internal temperature of $T_{int} = 100$ K.

The outcome of the HELIOS atmospheric calculations, also includes the corresponding emission spectra. Together with a PHOENIX spectrum (Husser et al. 2013) for the host star GJ 436, we calculate the secondary eclipse depths in the various Spitzer bandpasses for every model.

### 4.3 Consistent versus non-consistent joint atmosphere-interior models

Before, we apply the random forest model to our computed grid, we first describe the general outcomes of the coupled interior-atmosphere models as a function metallicity and carbon-to-oxygen ratio.

The example discussed in Section 3.3 corresponds to the situation where a physically consistent joint atmosphere-interior model could be found. This means that for the assumed atmospheric composition, and the associated resulting temperature and density at the BOA given by the atmosphere model, an interior structure could be found that also reproduces the observed total mass and radius.

As explained in Section 3.2.3, such a self-consistent coupled solution might, however, not exist for all assumed atmospheric compositions. For very strongly enriched compositions, the (mean) density of the envelope gas can alone (even in the absence of a dense solid core) become so high that it exceeds the observed mean density. In this case, no self-consistent/physical coupled model can be found, putting additional constraints on the allowed atmospheric compositions independent from the spectrum.

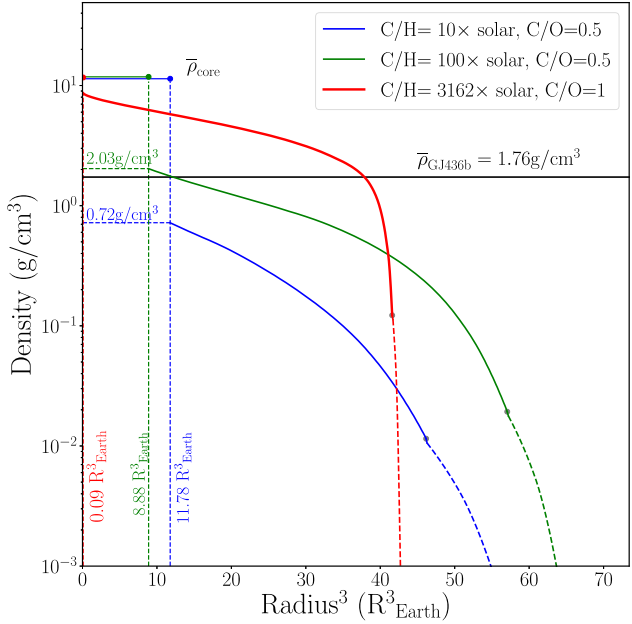

**Figure 5.** Density as a function of radius to the third power. We use the *x*-axis $R_p^3$ as it is a quantity proportional to the volume of the planet. The blue and the green lines correspond to cases with a physical joint atmosphere-interior solution of GJ 436 b whereas the red case does not. Although not shown (because of the very low associated densities), the green and the blue lines go to the same observed total radius $R = 4.19 R_\oplus$. The red line corresponding to a non-physical case does not, reflecting that for this strongly enriched case no self-consistent joint solution exists. The grey circle on top of each curve denotes the transition from the interior to the atmosphere (dashed line).

We demonstrate this with Fig. 5. It shows the envelope and mean core density for two cases (blue and green) for which we find a physical (self-consistent) solution and one for which no such solution exists (red). We plot in the *x*-axis the radius $R_p^3$ as it is proportional to the volume. The green line is the same case as studied in the previous section with C/H=100 × solar and C/O = 0.5. The blue line corresponds to a less enriched case with C/H=10 × solar and C/O = 0.5. For this case, the core and envelope mass fraction are 23.81 and 0.16 $M_\oplus$, respectively. The envelope mass fraction is thus lower than in the green case, while the core mass fraction is correspondingly higher. This can be seen by the higher $R^3$ at the core-envelope interface in the blue relative to the green case. In the plot, we also give the envelope gas density at the core-envelope boundary. It is 0.72 and 2.03 g cm$^{-3}$ for the C/H=10 × solar and C/H = 100 × solar compositions, respectively. The density is, as expected higher for the higher enrichment level. The observed mean density is in comparison 1.76 g cm$^{-3}$. To obtain it, the total mass must be distributed in an appropriate way into envelope (of lower density) and core (of higher density) such that the observed mean density result. For the C/H = 10 × solar case, the gas has a lower density than for the C/H=100 × solar case. Therefore, a higher core mass (i.e. more material at high density) is necessary to lead in combination to the observed mean density (again relative to the C/H=100 × solar case).

For the case of C/H = 3162 × solar and C/O = 1.0 (red line), which corresponds to a very dense gas, we did in contrast not find a solution. This is because for such a heavy-element dominated composition, the density in the atmosphere increases so fast with depth, that most of the planet volume is filled with envelope material that has a density higher than the observed mean one.

Given an envelope of very high density, the algorithm that tries to find the envelope mass fraction fitting the observed mean density decreases more and more the core mass fraction. But even for a planet made of solely envelope and atmosphere without core, the gas is already alone too dense relative to the observed value. Thus, no self-consistent coupled atmosphere-interior solution exists.

The red line is thus the last structure obtained in the iteration before the iteration stops because it is found that even for no core, the planet mean density is too high. We see that the radius is only about $43^{1/3} \approx 3.5 R_\oplus$, clearly falling short of the observed value of about $4.2 R_\oplus$. This extremely enriched atmospheric composition is thus inconsistent with the observed bulk density.

One of the motivations for the helium-atmosphere model in Hu et al. (2015) was that such high metallicities (1000 × or 10000 × solar) fail to match the bulk structure (planet's mass and radius) of GJ 436b. In this paper, we are also able to rule out such a dense atmosphere on the grounds of matching the planetary radius, but with the novel atmosphere-interior coupled models.

We also give the core mean density $\overline{\rho}_{\rm core}$ for all three cases. It is seen that the more massive the envelope, the higher the core density. The heavier envelope compresses the core more strongly through its weight, therefore increasing the density (e.g. Mordasini et al. 2012c).

### 4.4 Grid of coupled atmosphere-interior models for GJ 436 b

We now generalize to the entire grid in C/H and C/O. This gives an overview of how the interior structure (namely the envelope mass fraction and envelope metallicity *Z*) depend on the atmospheric composition. Importantly, it also shows which atmospheric compositions allow consistent coupled interior-atmosphere structures.

#### 4.4.1 Molecular weight $\mu$ as a function of metallicity and C/O

The upper left-hand panel of Fig. 6 shows a matrix built with different values of C/O and log (C/H) colour-coded according to the mean molecular weight $\mu$ of the gas in the envelope at the BOA, as predicted by HELIOS.

We find that at a fixed C/O value, increasing the metallicity C/H results in a linear (monotonically) increase of the molecular weight. For a solar-like composition (log(C/H) = 1), the values for $\mu$ are around 2.4, as expected for an $H_2$/He-dominated composition. At high metallicities of 100 or 316 × solar, clearly higher $\mu$ of 4 to 11 result, as expected from the chemical compositions shown in the pie charts in the third row of Fig. 1. There it can be seen how heavier molecules like $H_2O$, CO, or $CO_2$ become increasingly abundant with increasing C/H.

Increasing the C/O ratio for a fixed C/H value leads to monotonically decreasing mean molecular weights. This is also connected to the chemical composition of the atmospheres. As explained in Section 2.2, at higher C/O values, $H_2O$ disappears, while the atmosphere starts to become more dominated by $H_2$ instead. We remind the reader that in our scheme of specifying atmospheric composition, the carbon abundance is given by the C/H value, while the oxygen is varied in a way to give the requested C/O. For the high C/O cases, most of the available oxygen is increasingly bound in the chemically very stable molecule CO, such that only little free oxygen atoms are available to form water. This implies that at C/O=1, there are 10 times less free oxygen atoms than in the C/O = 0.1 case. This can directly be seen, for example, in the lowest panel of Fig. 1, where increasing the C/O ratio from 0.1 to 1 leads to large increase in the abundance of $H_2$ and (except for the C/H=10000 × solar case) to a strong decrease of $H_2O$. This effect is also clearly evident in the bottom right-hand panel of Fig. 6.

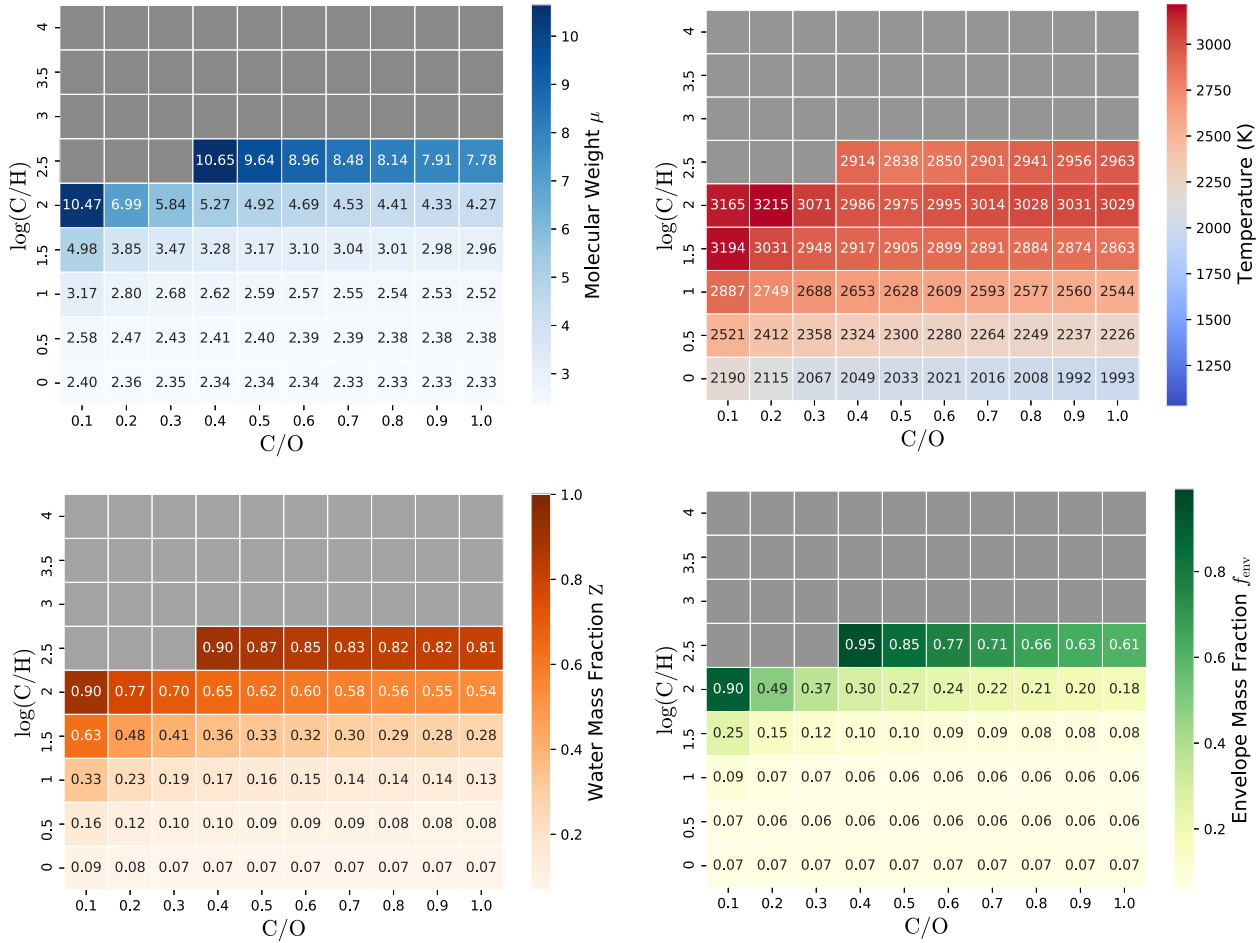

**Figure 6.** Grid of coupled atmosphere-interior structures for GJ 436 b. The four panels show as a function of C/H and C/O four quantities characterizing the interior structure. The top left panel is the mean molecular mass $\mu$ at the BOA (i.e. coupling point of atmosphere and interior), which is an output of HELIOS. The top right-hand panel is the temperature at the BOA, which is again a output of HELIOS, and one of the outer boundary conditions for the interior structure model. The bottom left-hand panel is the water mass fraction $Z$ used in the EOS of the interior structure determined by requiring that the density is continuous at the BOA. Gray fields are C/H–C/O pairs that do not allow a consistent atmosphere – interior coupling, because the enrichment in heavy elements and thus gas density is too high to reproduce the observed mean density, even in the complete absence of a solid core. The bottom right-hand panel is finally the envelope mass fraction as key output of the interior structure model, the rest being the silicate-iron core.

As stated in Section 3.1, the maximum possible $\mu$ in the EOS of interior structure model is 18, which corresponds to the mean molecular weight of (pure) water. Higher $\mu$ values predicted by the atmospheric model would require that the density in the interior structure is scaled. Fortunately, we see that for all converged models, the $\mu$ remained smaller, meaning that at least for these models, no scaling factor was necessary.

*4.4.2 Temperature at the BOA as a function of metallicity and C/O*

The temperature at the bottom of the atmosphere is shown in the upper right-hand panel of Fig. 6. The distribution of these temperatures does not always show the same overall trends as those of the discussed quantities in this section. The temperature depends strongly on the actual temperature–pressure profile of the atmosphere. This profile, however, is not only determined by the bulk of the atmosphere but usually by smaller trace gases (e.g. $CH_4$) that, while having a rather small abundance, interact strongly with the radiation field via absorption and emission. Thus, the atmospheric temperature profile is more determined by the abundances of these trace gases that do not impact other bulk quantities of the atmosphere and envelope, such as the mean molecular weight, to a large degree. However, we can observe some overall, global trends. With increasing C/H values, the temperatures do also increase. This is caused by higher abundances of strongly absorbing molecules.

The $\mu$ and temperature at the BOA shown in the top row are outputs of the atmospheric model HELIOS, and inputs to the interior structure model Completo21. In the bottom row, we show to which values they lead for two quantities characterizing the interior (and used by the interior model to calculate the interior structure), namely the water mass fraction $Z$ in the EOS of the interior, and most importantly, the envelope mass fraction.

*4.4.3 Water mass fraction Z as a function of metallicity and C/O*

The bottom left-hand panel depicts the dependence of the water mass fraction $Z$, with C/O and C/H in the envelope. $Z$ is a placeholder for all heavy elements, and it is obtained by requiring that the density obtained with the interior structure's EOS is the same as the one in the atmospheric model. The observed behaviour of $Z$ is directly connected to the chemical composition and, thus, its behaviour as a function of C/O and C/H is similar to that of $\mu$. We, however, observe the following: for a $\mu$ of about 2.4, one would expect an associated $Z$ of about 0.01 to 0.02, the solar value. However, the $Z$ that is actually found is rather 0.07 to 0.09. This is a consequence of requiring that the density is continuous, and not $\mu$, and the fact that one EOS is ideal, and the other not. None the less, the general pattern is clear: low C/H lead to a low $Z$, i.e. a H/He-dominated interior, whereas high C/H correspond to a metal-rich interior. The highest $Z$ is 0.9, meaning that the interior of this planet is strongly dominated by species different than H/He (in the interior EOS represented by water).

#### 4.4.4 Envelope mass fraction as a function of metallicity and C/O

As the most important output of the interior structure model, the bottom right-hand panel shows the envelope mass fraction values $f_{env}$ for each case as a function of C/O and log (C/H). The figure shows that with increasing metallicity and decreasing C/O ratio, the envelope mass fraction increases. This is caused by the chemical composition being more and more dominated by heavier molecules, such as CO or $H_2O$, at high metallicity values, rather than the lighter $H_2$ and He species at metallicity values resembling more solar elemental abundances (see also Fig. 1). This means that the density of the envelope gas increases, which means that a less massive silicate-iron core (of even higher density) is needed to obtain the observed mean density of 1.76 g cm$^{-3}$, as discussed in Section 4.3.

At log (C/H) = 2.5 and C/O = 0.4, for example, the envelope mass fraction is 0.95 (containing large amounts of species like $H_2O$ or CO), meaning that silicate-iron core is only 5 per cent of the total mass. Such an envelope mass that is much more massive than the core mass is rather unlikely from a planet formation and bulk composition or disc chemistry point of view. For a formation beyond the water iceline, a volatile fraction of about 0.5 is expected (e.g. Lodders 2003). Various processes like the accretion of dry material inside the iceline (Figueira et al. 2009) or building-block devolatilzation (Lichtenberg & Krijt 2021) would further reduce this fraction. Such a very low-mass core could also not lead to the outgassing of such a massive envelope (Elkins-Tanton & Seager 2008). Thus, while formally allowed, we think that values of $f_{env} \geq 0.5$ are rather unlikely.

At a solar C/H, the envelope mass fraction, which is in this case strongly H/He dominated, is 7 per cent. A $\sim$10 per cent H/He mass fraction in GJ 436 b is in agreement with past work (e.g. Adams, Seager & Elkins-Tanton 2008).

We also observe that when we require that the envelope mass fraction is less than $\sim$50 per cent, the maximum $\mu$ in the valid atmosphere is $\sim$4 corresponding to log (C/H) = 2. The exact value of $\mu$, however, also depends on C/O. This upper bound for $\mu$ is interesting, as $\mu$ is important to calculate the scale height which in turn is important to determine the shape of the spectrum.

Finally, beyond some high values for C/H (100 $\times$ solar for low C/O and 315 $\times$ solar for higher C/O) we are unable to find a converged, consistent solution for both `Completo21` and `HELIOS` (see also Section 3.1). These cases are colored in grey in all four different panels. For these excluded cases, the envelope gas is so dense that even in absence of a (even denser) core, the observed rather low mean density cannot be reproduced.

The plot can, however, also been read in another way: namely that the bulk density can be reproduced with a wide range of degenerate compositions, ranging from silicate/iron planets with about 10 per cent H/He to planets consisting almost entirely of water (or more generally speaking, volatiles other than H/He). This degeneracy in the mass–radius relation of (sub)Neptunian planets was pointed out by Adams et al. (2008), who also speculated that spectroscopy will be the most likely means to break the degeneracy.

**Table 2.** *Spitzer* transmission and secondary-eclipse measurements based on Lanotte et al. (2014).

| Bandpass (μm) | Depth (ppm) |
| --- | --- |
| Transit | |
| 3.6 | 6770 $\pm$ 42 |
| 4.5 | 6881 $\pm$ 54 |
| 8 | 6789 $\pm$ 61 |
| Secondary eclipse | |
| 3.6 | 177 $\pm$ 43 |
| 4.5 | 28 $\pm$ 25 |
| 5.8 | 229 $\pm$ 107 |
| 8.0 | 362 $\pm$ 29 |
| 16 | 1260 $\pm$ 280 |
| 24 | 1690 $\pm$ 470 |

### 4.5 Retrieval analysis of GJ 436 b

Using the grid, we now apply `HELA` to the *Spitzer* secondary eclipse observations from Table 2. We obtain posterior distributions for C/H and C/O.

The right-hand panel of Fig. 7 shows the posterior distributions of C/O and log (C/H). We observe that a C/O $\sim$ 1 and a very high metallicity ($\sim$ 3162$\times$ solar) are needed to match the *Spitzer* observations by Lanotte et al. (2014). However, a metallicity $\geq$ 1000$\times$ solar is likely unphysical. This is shown by the first panel in Fig. 6, where at such high metallicities the gas alone in the interior is so dense that we cannot reproduce the bulk density of GJ 436 b, even in the absence of a core (see Fig. 5). Thus, the assumptions underlying the retrieval analysis lead to a derived atmospheric composition which is excluded based on the independent interior structure constraints arising from the observed bulk mass and radius. This in turn indicates that some of these assumptions must be revised.

The left-hand panel of Fig. 7 shows the median-fit of GJ 436 b for the emission spectra. Similar to other studies (Madhusudhan & Seager 2011; Lanotte et al. 2014; Morley et al. 2017) our model fails to fit the 4.5-μm *Spitzer* IRAC channel. The main reason for this is the underlying assumption of chemical equilibrium, which we have adopted in this study. In chemical equilibrium, and at an equilibrium temperature of $\sim$ 1000K, we expect $CH_4$ to be the most abundant carbon carrier but not CO. Methane absorbs strongly in the 3.6-μm band and CO and $CO_2$ absorb in the 4.5-μm band. However, the 4.5-μm point exhibits high absorption from CO and or $CO_2$, more than can be explained with chemical equilibrium. The abundance of CO is tied closely to the abundance of $CO_2$ since they both have strong features at 4.5 μm. Even if a high $CO_2$ can be explained solely on the grounds of equilibrium chemistry and high metallicity (Madhusudhan & Seager 2011), the high CO requires both high metallicity and non-equilibrium chemistry. Low abundance of $CH_4$ could also be explained with disequilibrium chemistry (Zahnle, Marley & Fortney 2009a).

Morley et al. (2017) also showed that this discrepancy is unlikely to be due to clouds and hazes. For completeness, we show in Fig. 8 the corresponding transmission spectra for the median-fit retrieved model in Fig. 7. We see that the transmission spectrum does not reproduce the 4.5-μm data point either. It is, thus, likely that the atmosphere of GJ 436 b is in chemical disequilibrium (Stevenson et al. 2010; Madhusudhan & Seager 2011; Morley et al. 2017). This chemical disequilibrium could be a byproduct of different effects like vertical mixing and photochemistry (Line et al. 2011).

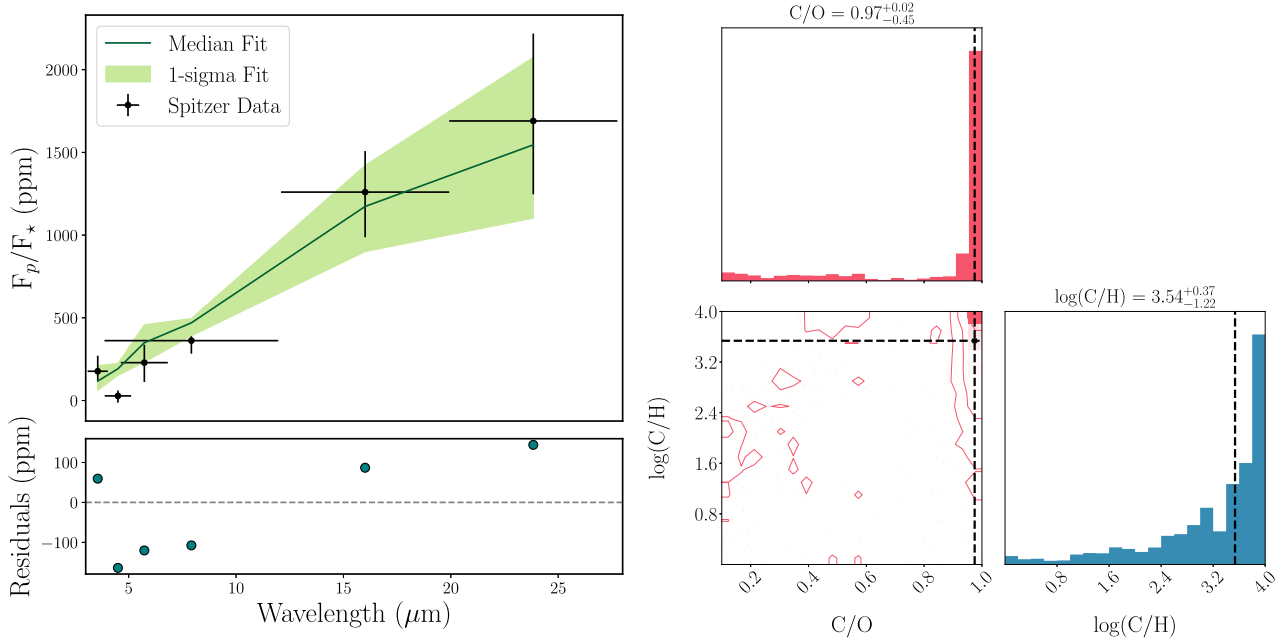

**Figure 7.** Atmospheric retrieval of the emission spectrum of GJ 436 b, which consists of the six *Spitzer Space Telescope* measurements reported by Lanotte et al. (2014). The retrieval uses the random forest method. Left-hand panels: data points, median-fit model, and residuals. Right-hand panels: posterior distributions of C/O and C/H. The vertical black dotted lines show the median value of each posterior distribution, which are also indicated numerically above each panel.

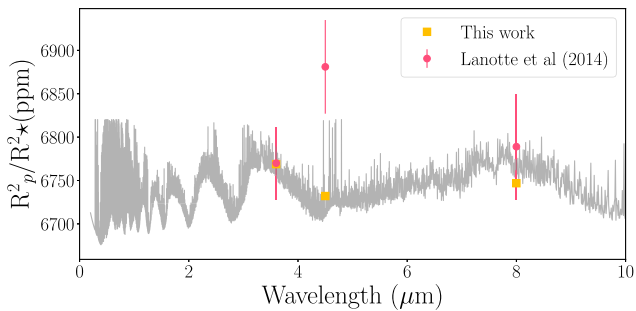

**Figure 8.** Transmission spectrum corresponding to the median-fit parameters in Fig. 7. Overplotted in pink are the three *Spitzer* transmission points from Lanotte et al. (2014) and in yellow squares our predicted values.

Given these insights, the discrepancy between retrieved atmospheric composition and the interior structure constrains is most likely explained by the assumption of chemical equilibrium in the atmospheric model.

## 5 DISCUSSION

### 5.1 Comparison to past work

#### 5.1.1 Interpretations of the emission spectra of GJ 436 b

Madhusudhan & Seager (2011) were the first study to point out the significance of the 4.5-μm data point in the emission spectrum of GJ 436 b. Fig. 4 of that study illustrates the difficulty of fitting this data point without invoking chemical equilibrium. Figs 22 and 23 of Lanotte et al. (2014) corroborate this finding. Madhusudhan &

Seager (2011) also suggested that a high abundance of $CO_2$ is needed to fit the emission spectrum, which results in a supersolar metallicity.

Morley et al. (2017) explored a variety of possible reasons for the discrepancy of the 4.5-μm data point with models. Fig. 7 of that study shows how even chemical disequilibrium models struggle to match the 4.5 μm data point. Fig. 11 of the same study shows that interior heating has a negligible effect on the 4.5-μm data point. Fig. 12 shows how the addition of clouds, as prescribed using the approach of Ackerman & Marley (2001), worsen the match to the 3.6-μm data point. Fig. 13 shows how hazy models are overall a poor match to the emission spectrum of GJ 436 b. Generally, Morley et al. (2017) do not report models that are able to fit all *Spitzer* data points.

#### 5.1.2 Interior model relative to Morley et al. (2017)

Our interior structure model differs in a number of aspects to the one used in Morley et al. (2017). First, Morley et al. (2017) assume an inert core composed of either 50 per cent rock and 50 per cent ice or 100 per cent water ice. In our model, the solid core is in contrast composed of silicates and iron in a 2:1 mass ratio. From a planet formation and condensation sequence point of view, it appears difficult to imagine that water alone condenses but not silicates and iron (for the 100 per cent water core case in Morley et al. 2017), or that silicates and water condense but not iron (for their 50 per cent rock and 50 per cent ice core case). Therefore, core compositions made of silicate and iron appear more likely, which is also supported by recent exoplanet observations (Adibekyan et al. 2021).

The consequence of these different assumed core compositions is that the cores in Morley et al. (2017) should likely have lower densities than in our model. This allows Morley et al. (2017) to accommodate more enriched envelopes while still reproducing the observed mass and radius (i.e. the mean density). In our model, the highest enrichment reproducing the observed mean density is $\log(C/H) = 2.5$ corresponding to about 316 × solar at C/O ≥ 0.4, or $\log(C/H) = 2$ (100 × solar) at lower C/O (see Fig. 6). In this plot, the grey cases are the excluded ones in our model, where the gas alone is so dense that we cannot reproduce the observed mean density of 1.73 g/cm$^3$, even in absence of a core.

A second difference is that we assume that water is always part of the envelope in the form of a homogeneous mixture of H/He and H$_2$O, using the additive volume law. In some models of Morley et al. (2017) water is in contrast a part of the inert core, and not mixed with the envelope. Given the high temperature in both the envelope and interior of GJ 436 b and the fact that water and H/He are miscible at typical interior temperatures and pressures (Soubiran & Militzer 2015), it appears likely that the water does not reside in a separate pure water layer, but that it is rather mixed with H/He. It is, however, unclear to what extent such a mixture would be homogeneous, since the presence of compositional gradients in the interior may inhibit or weaken convective mixing (Nettelmann et al. 2013). Such gradients could, in turn, be the consequence of the accretion history of the planet (Valleta & Ravit 2020).

Third, Morley et al. (2017) use the ANEOS equations of state from Thompson (1990) for water and water-rock and Saumon, Chabrier & van Horn (1995) (SCvH) for H/He. In our model, we use for water the AQUA EOS from Haldemann et al. (2020), which, in turn, includes the Matzevet et al. (2019) water EOS at the high densities. For H/He we use the new CMS EOS from Chabrier et al. (2019). This new generation of EOSs are based on ab initio first principles quantum molecular dynamics simulation and should be more realistic than the older semi-analytical approaches. This is indicated by them reproducing experimental data clearly better (Chabrier et al. 2019).

Fourth, in Morley et al. (2017) the pure water and water-rock equations of state are independent of temperature. In our case, the temperature dependence of water is fully included via the AQUA EOS. For the silicate-iron cores, we approximately take into account their thermal expansion, as described in Linder et al. (2019).

On the other hand, we model the metals mixed into the H/He as water only, while Morley et al. (2017) also consider mixtures of water + rock + H/He. We can therefore not explore in details the consequences of a possible accretion and subsequent enrichment that is dominated by rocky rather than icy materials, which could be important for GJ436 b (Figueira et al. 2009). We also did not explore the consequences of different interior temperatures (Morley et al. 2017). But independently of these limitations and differences, both models find that very high atmospheric enrichment levels are inconsistent with the bulk interior, and that for a better understanding of the planet, a self-consistently coupled approach for the atmosphere and interior should be used.

## 5.2 Implications for future observations and their interpretations with *JWST*

The random forest method from the previous retrieval analysis also yields additional important outcomes that can be used to estimate the possibility for future instruments to provide better atmospheric characterization than available with the limited *Spitzer* and WFC3 data.

One important result of the random forest are the feature importance and the real-versus-predicted plots, which quantify the degree to which each parameter may be predicted in mock retrievals given a noise model. This provides crucial information on the wavelength ranges that should be observed for the most optimum atmospheric characterization given a certain noise level.

Figs 9 and 10 show the outcomes of performing 10 000 mock retrievals on GJ 436 b emission and transmission spectra, respectively. We consider both noisy and noise-free models for each case.

The ability of the random forest to accurately predict the outcome is quantified by the coefficient of determination $\mathcal{R}^2$.

For the transmission spectra, we use a wavelength range from 0.2 to 5.3 μm representative of the coverage of *JWST* instruments to measure transmission spectra of exoplanet atmospheres, like the *Near InfraRed Spectrograph* (NIRSpec) (0.6–5 μm) and the *Near-Infrared Imager and Slitless Spectrograph* (NIRISS) (0.6–5μm) (Greene et al. 2016). For emission spectra, on the other hand, we consider the wavelength range from 0.3 to 20 μm. This is part of the range covered by *JWST*'s *Mid-Infrared Instrument* (*MIRI*) (5–28 μm) and NIRSpec (0.6–5 μm).

Our results shown in the top left-hand panel of Figs 9 and 10 suggest that both emission and transmission spectra do equally well at predicting the C/O ratio and metallicity in the considered wavelength ranges. Even in the presence of noise, the retrieval of both quantities is still robust. Further increasing the noise to 20 or 100 ppm does not decrease the predictive power significantly

For completeness, the feature importance plots for each case are also depicted in the top right-, bottom right- and bottom left-hand panels of Figs 9 and 10. These plots show the fractional importance of each data point for constraining the C/O ratio and overall metallicity C/H.

For transmission spectra, 2.8 and 5.4 μm seems to be the most important for constraining the metallicity and C/O ratio. This is caused by the presence of major absorption bands of important carbon and oxygen-bearing molecules, such as $CH_4$ (3.1–3.9 μm), CO (3.9–5.1 μm), and $CO_2$ (4–5 μm). Since the infrared wavelength region contains multiple absorption bands of most major molecules, the emission spectrum at these wavelengths easily allows us to predict the metallicity and the C/O ratios as suggested by the feature importance plots in Fig. 9. Adding noise to either the transmission or emission spectra mutes some of the weak features at shorter wavelengths.

As an additional pathway to atmospheric diversity, Hu et al. (2015) suggested that fractionation during the long-term atmospheric escape from sub-Neptunes may lead to the depletion of hydrogen but the retention of helium, which has the effect of enriching the elemental abundances of carbon, oxygen and nitrogen relative to hydrogen. Together with the findings of the current work, this suggests that the atmospheric chemistry of objects in the sub-Neptunian regime and across the Fulton gap (also known as the radius valley, Fulton et al. 2017) possesses a richness and diversity far beyond what is expected in hot and warm Jupiters. The *JWST* is well-equipped to accurately measure the abundances of the major carbon, oxygen and nitrogen carriers (Guzmán-Mesa et al. 2020). This should, for example, allow to derive additional observational constraints on the various hypotheses for the origin of the radius valley. Some of these hypotheses predict that the planets above the valley should be rocky with a per cent-level H/He envelope (e.g. Owen & Wu 2017; Jin & Mordasini 2018), while others rather predict a water-rich composition with an ∼50 per cent water mass fraction (Venturini et al. 2020a). These different compositions put strong constraints on formation models regarding the extent of orbital migration (Baruteau et al. 2016).

## 5.3 Future model grids without chemical equilibrium

In this study, we have demonstrated that joint atmosphere-interior models are required to properly interpret emission spectra of GJ 436 b – and sub-Neptunes in general. However, we have constructed atmospheric model grids that assume chemical equilibrium. A natural generalization of the current work is to relax this assumption. The simplest approach to chemical disequilibrium is the 'quenching approximation' (Prinn & Barshay 1977), which asserts that there is a point (pressure level) in the atmosphere where the chemical and dynamical time-scales are equal. Above this 'quench point', the abundances of chemical species are frozen at their chemical equilibrium value at the quench point. Madhusudhan & Seager (2011) implemented the quenching approximation with a single quench point for all molecules considered in their model. An outstanding challenge is how to adequately sample the multidimensional parameter space of a model grid with ∼10 or more parameters. A recent study demonstrated that the optimal strategy is to sample randomly; this approach appears to produce comparable outcomes to Latin-hypercube sampling (Fisher & Heng, in preparation).

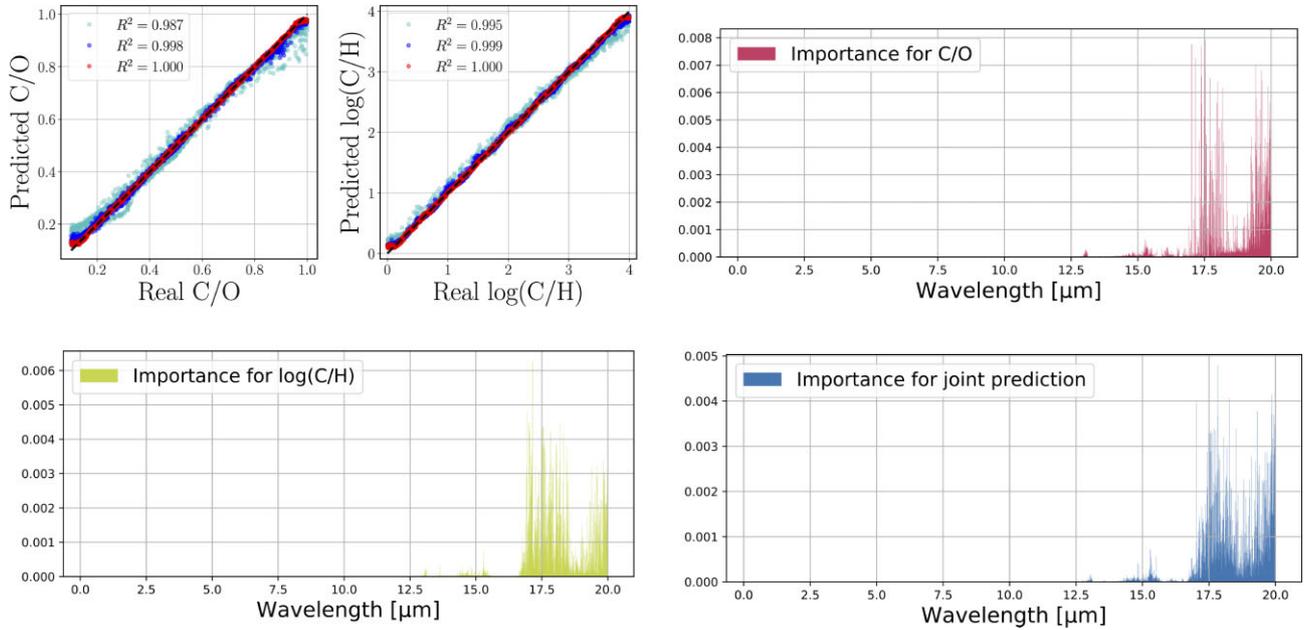

**Figure 9.** Real versus predicted (RvP) values and feature importance of C/O and C/H using the supervised machine learning method of the random forest. From the training set of 10000 synthetic emission spectra (0.34–20 µm), we use 8000 models for training and 2000 models for testing; the RvP and feature importance plots are the outcome of this testing step. The stellar and exoplanetary parameters of GJ 436 and the warm Neptune GJ 436 b, respectively, are assumed (see the text). The top left-hand panel panels are the RvP calculations. We consider three different different noise scenarios: noise-free models (red), an optimistic photon-limited uncertainty of 20 parts per million (ppm) (blue) and 100 ppm photon-noise (cyan). Negative and positive values of the coefficient of determination ($-1 \leq \mathcal{R} \leq 1$) correspond to negative and positive correlations, respectively. The remaining three panels (top left-hand panel, bottom left-hand panel, and bottom right-hand panel) are the feature importance calculations quantifying the information content of emission spectra as functions of wavelength. All of the entries in each feature importance plot add up to unity. Shows in the feature importance for the 100 ppm case.

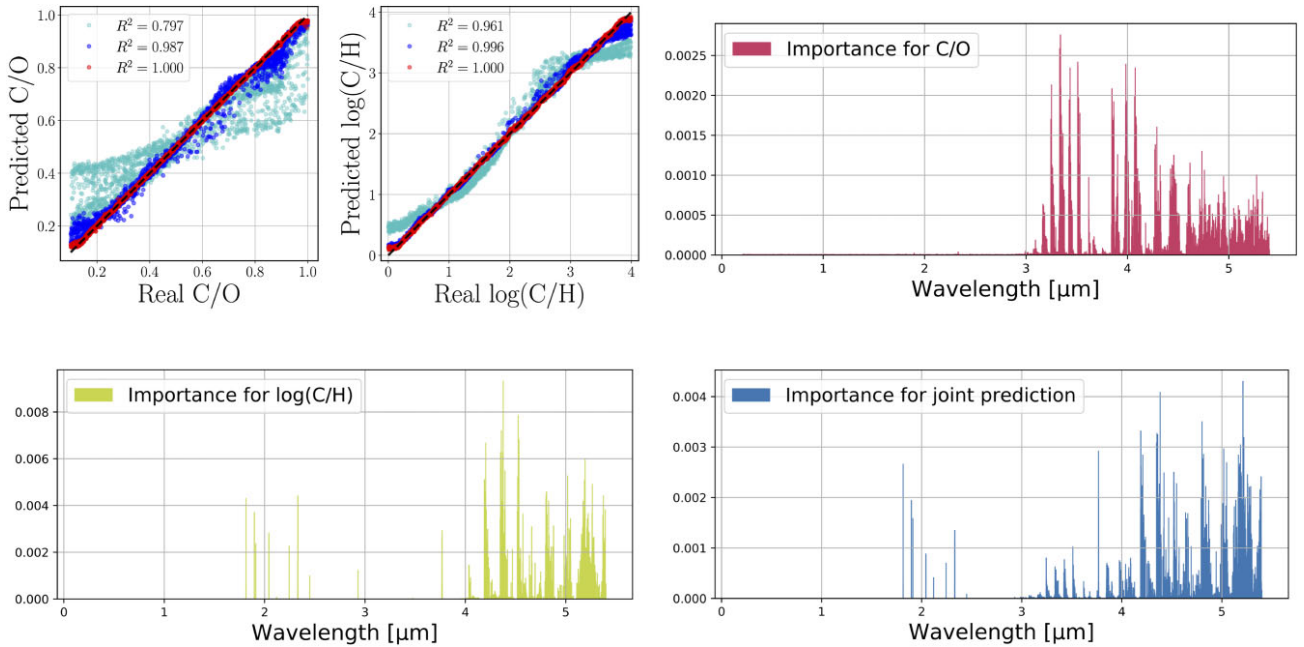

**Figure 10.** Same as Fig. 9, but for transmission spectra.

## 6 SUMMARY

We have confirmed the atmospheric diversity of sub-Neptune atmospheres presented in Moses et al. (2013), and also predicted by Hu & Seager (2014), using self-consistent radiative transfer and equilibrium chemistry models. We coupled the atmosphere and interior structures of GJ 436 b finding that very high-metallicity atmospheric models ($\geq 1000 \times$ solar) are excluded by the interior structure. We apply the random forest machine learning technique to available *Spitzer* data finding that even though is possible to fit the emission spectrum with a high metallicity and a high carbon-to-oxygen model, such an atmosphere is inconsistent with physically plausible interior structures. The previous results suggest that GJ 436 b's atmosphere could be in chemical disequilibrium. In view of the recently launched *JWST*, we recommend that emission and transmission spectra of sub-Neptune planets are analyzed in a self-consistent coupled way using both the atmospheric and interior structures.

## ACKNOWLEDGEMENTS

We acknowledge financial support from the Swiss National Science Foundation, the European Research Council (via a Consolidator Grant to KH; grant number 771620), the PlanetS National Center of Competence in Research NCCR PlanetS, the Center for Space and Habitability (CSH) and the Swiss-based MERAC Foundation. CM acknowledges the funding from the Swiss National Science Foundation under grant 200021_204847 'PlanetsInTime'. We are grateful to Brett Morris and Chloe Fisher for constructive discussions and advice on the manuscript. We thank the referees for the constructive feedback that helped to improve this work.

## DATA AVAILABILITY

No new data were generated or analysed in support of this research.

# APPENDIX A: FULL CHEMICAL AND TEMPERATURE–PRESSURE PROFILES

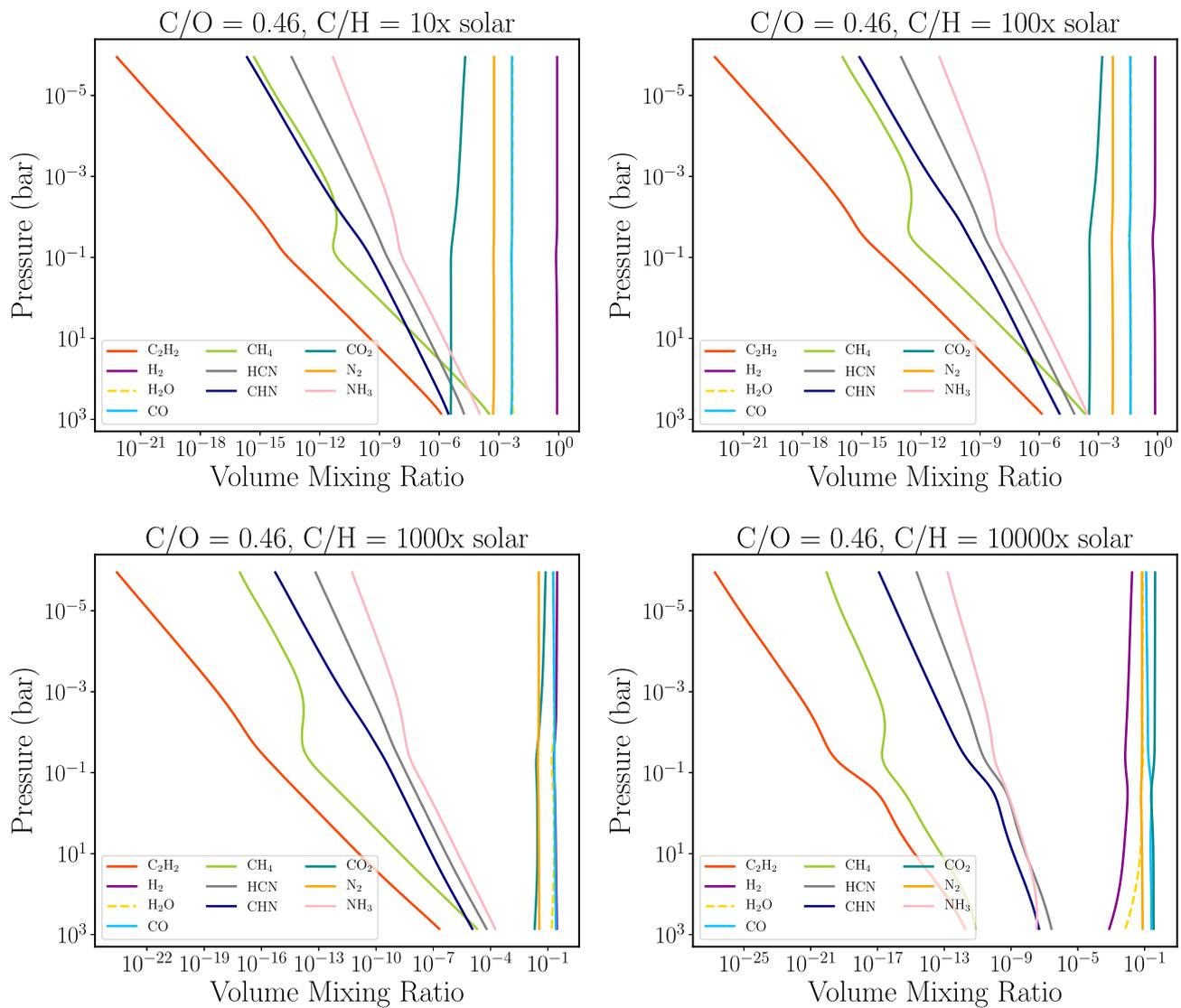

**Figure A1.** Full profiles of chemical abundances by number (volume-mixing ratios) as functions of pressure, corresponding to the 'chemical pie charts' shown in Fig. 1. The corresponding temperature–pressure profiles are shown in Fig. A5. This case corresponds to a fixed C/O=0.46.

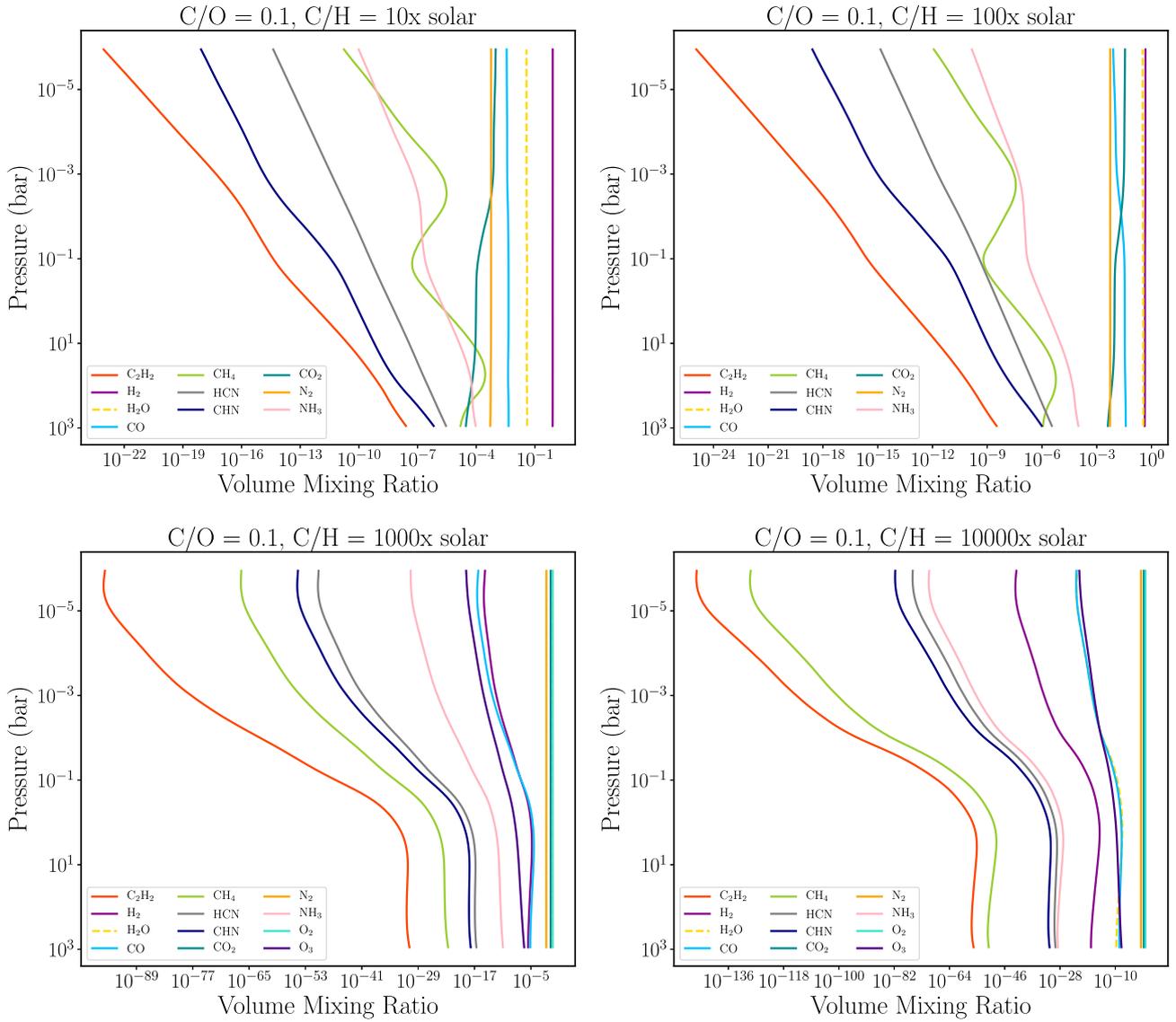

**Figure A2.** Same as Fig. A1 but for a fixed C/O=0.1.

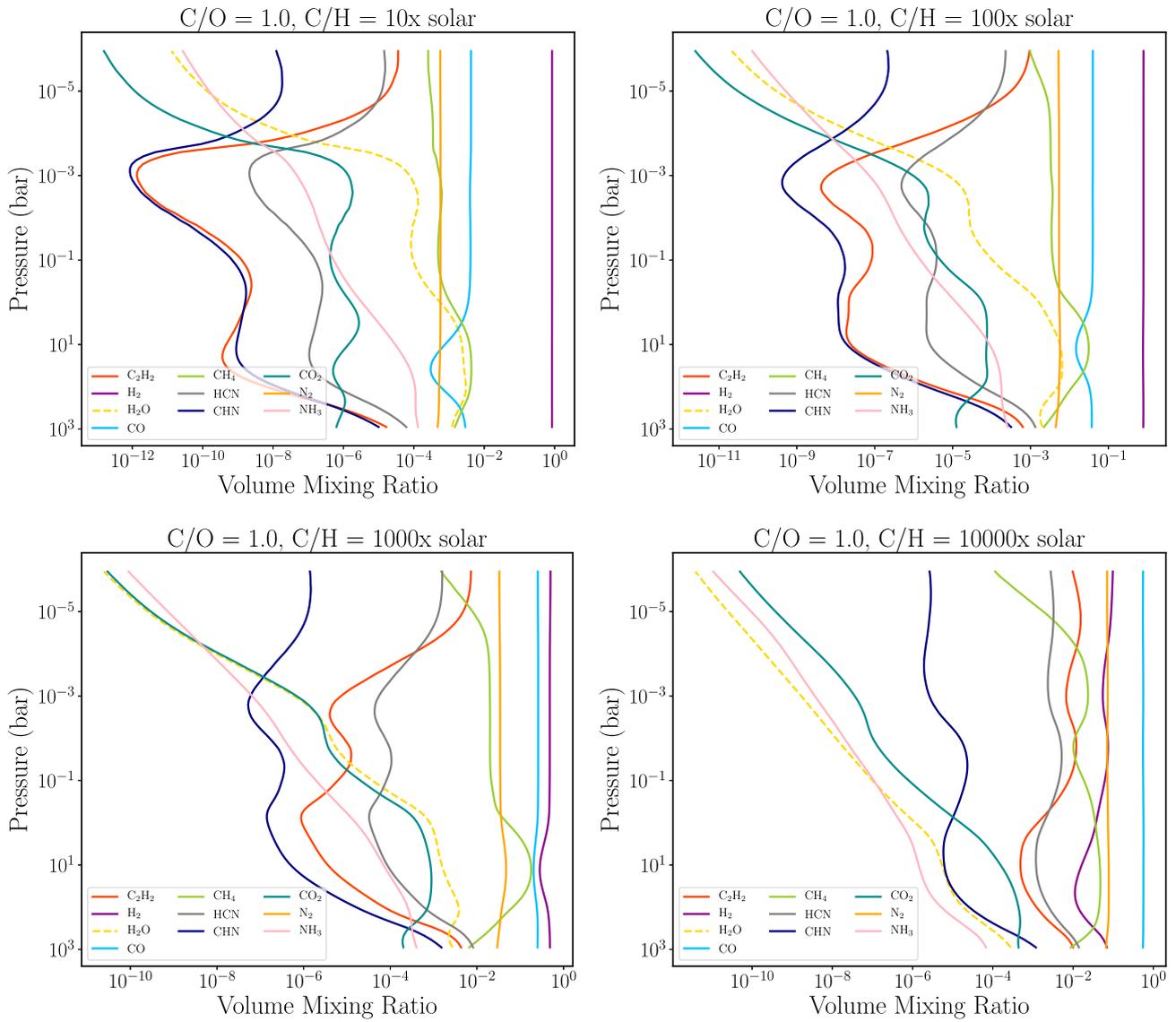

**Figure A3.** Same as Fig. A1 but for C/O = 1.0.

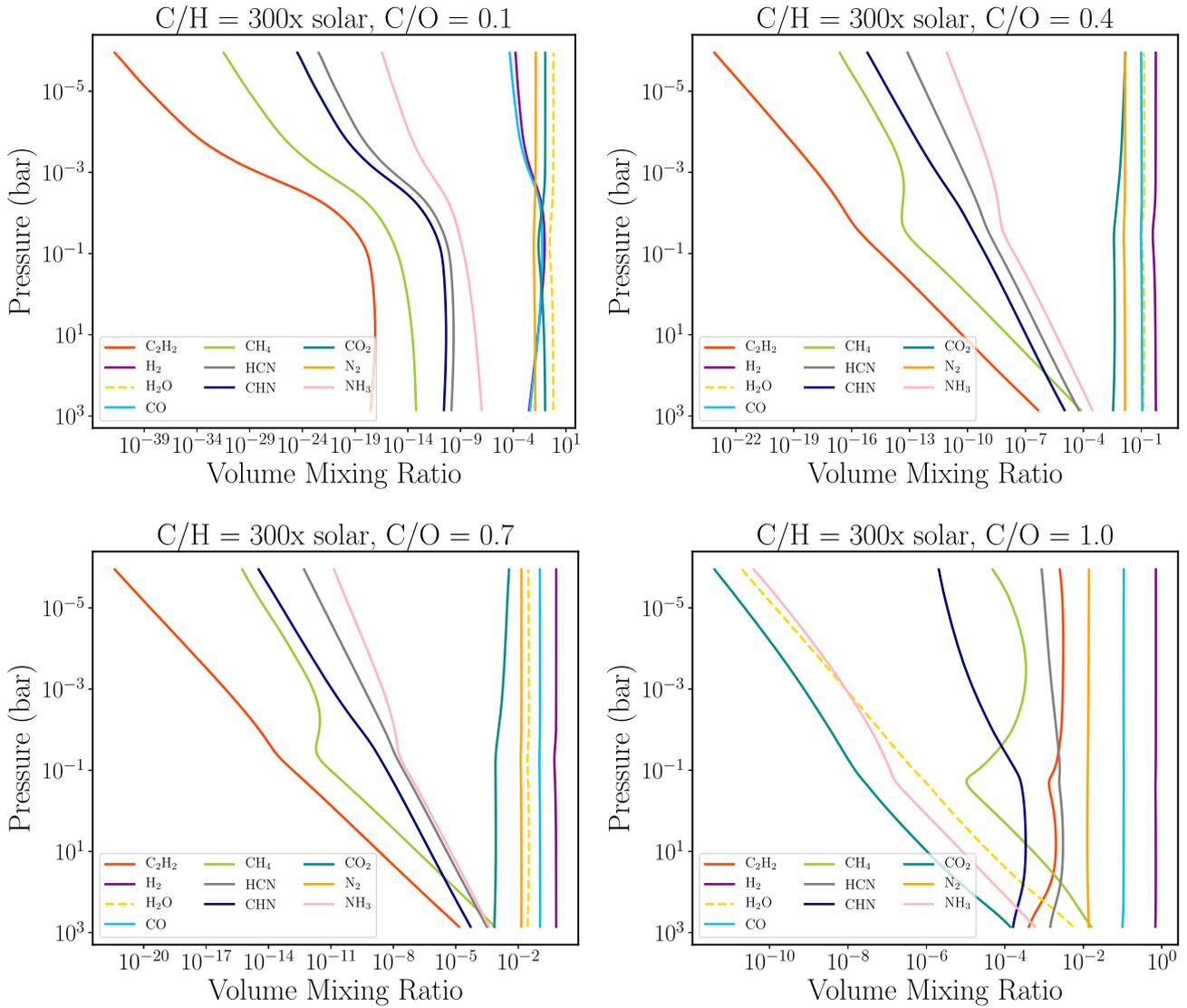

**Figure A4.** Same as Fig. A1 but for a fixed C/H = 300 × solar.

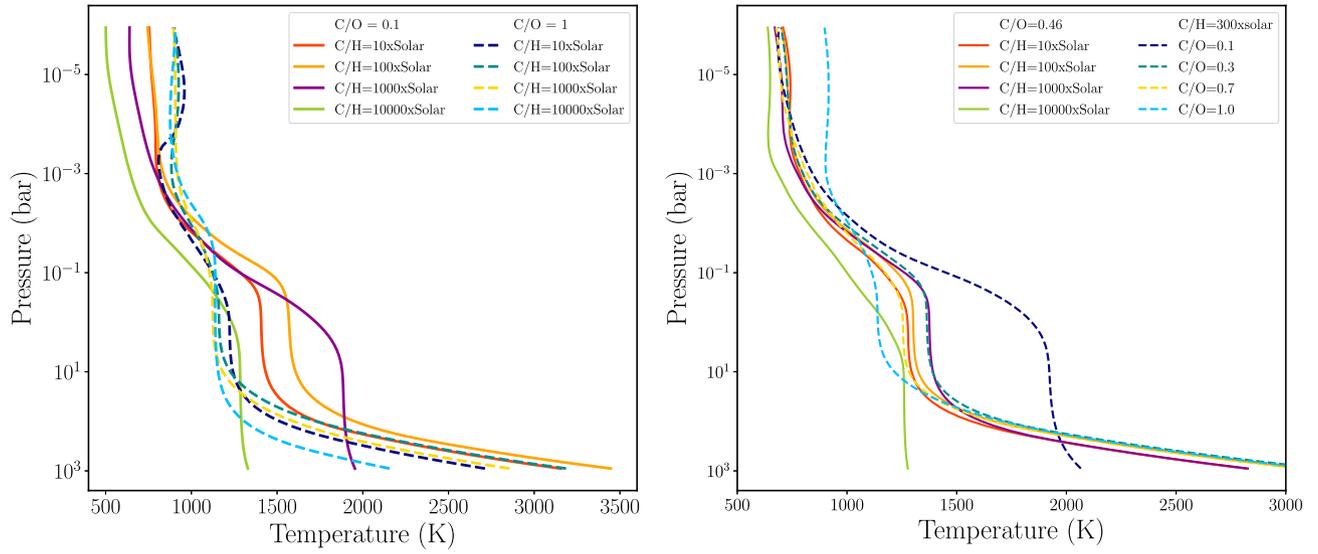

**Figure A5.** Temperature–pressure profiles of self-consistent radiative transfer and equilibrium chemistry models corresponding to Fig. 1. For C/O = 0.1 and C/H = $10^4 \times$ solar, the diminished temperatures are due to missing opacity sources for molecular oxygen.

This paper has been typeset from a T$_{\rm E}$X/L$^{\rm A}$T$_{\rm E}$X file prepared by the author.